\documentclass[12pt]{article}
\usepackage[utf8]{inputenc}
\usepackage{dsfont}
\usepackage{lineno}
\usepackage{setspace}
\usepackage{soul}
\usepackage{amsmath}
\usepackage{amsfonts}
\usepackage{titling}
\usepackage{amssymb}
\usepackage{mathtools}
\usepackage[authoryear]{natbib}
\usepackage{caption} 
\usepackage[margin=1in]{geometry}
\usepackage{titlesec}
\titlelabel{\thetitle.\quad}
\titleformat{\subsection}    
       {\large\slshape\bfseries}{\thesubsection.}{1em}{}
\titleformat{\subsubsection}
   {\slshape\large}{\thesubsubsection.}{1em}{}

\title{Accounting for Location Measurement Error in Imaging Data with Application to Atomic Resolution Images of Crystalline Materials}
\author{Matthew J. Miller$^a$, Matthew J. Cabral$^b$, Elizabeth C. Dickey$^b$, \\James M. LeBeau$^c$,  and Brian J. Reich$^a$}
\date{}

\begin{document}
\maketitle
\begin{singlespace}
\vspace{-.5in}
\begin{enumerate}
\item[$^a$]  Department of Statistics, North Carolina State University, Raleigh, NC
\item[$^b$]  Department of Materials Science and Engineering, North Carolina State University Raleigh, NC
\item[$^c$]  Department of Materials Science and Engineering, Massachusetts Institute of Technology, Cambridge, MA
\end{enumerate}
\end{singlespace}
\noindent \textbf{Key words}: spatial statistics, Bayesian hierarchical modeling, scanning transmission electron microscopy (STEM), image analysis, materials science
\doublespacing
%\linenumbers
\begin{abstract}
\begin{singlespace}
\noindent Scientists use imaging to identify objects of interest and infer properties of these objects. The locations of these objects are often measured with error, which when ignored leads to biased parameter estimates and inflated variance. Current measurement error methods require an estimate or knowledge of the measurement error variance to correct these estimates, which may not be available. Instead, we create a spatial Bayesian hierarchical model that treats the locations as parameters, it using the image itself to incorporate positional uncertainty. We lower the computational burden by approximating the likelihood using a non-contiguous block design around the object locations.  We apply this model in a materials science setting to study the relationship between the chemistry and displacement of hundreds of atom columns in crystal structures directly imaged via scanning transmission electron microscopy. Greater knowledge of this relationship can lead to engineering materials with improved properties of interest. We find strong evidence of a negative relationship between atom column displacement and the intensity of neighboring atom columns, which is related to the local chemistry. A simulation study shows our method corrects the bias in the parameter of interest and drastically improves coverage in high noise scenarios compared to non-measurement error models.
\end{singlespace}
\end{abstract}
\section{Introduction}

\begin{figure}
 \centering
  \caption{\textit{Left}: Scanning Transmission Electron Microscope (STEM) image of Lead Magnesium Niobate (PMN), with red boxes placed around the A-site columns and blue boxes around the B-sites columns. \textit{Bottom Right}: Zoomed in view of atom columns with plotted centers of the columns found from nonlinear least squares, where $\hat{\mathbf{s}}_{Bj}$ and $\hat{\mathbf{s}}_{Ak}$ are the estimated locations of the $j^{th}$ B-site column and $k^{th}$ A-site column, respectively. \textit{Top Right}: Rendering of the crystal structure of PMN, showing the A-sites as columns of lead (gray) and the B-sites as columns of niobium (green) and magnesium (yellow).}
 \includegraphics[page=1,width=1\textwidth]{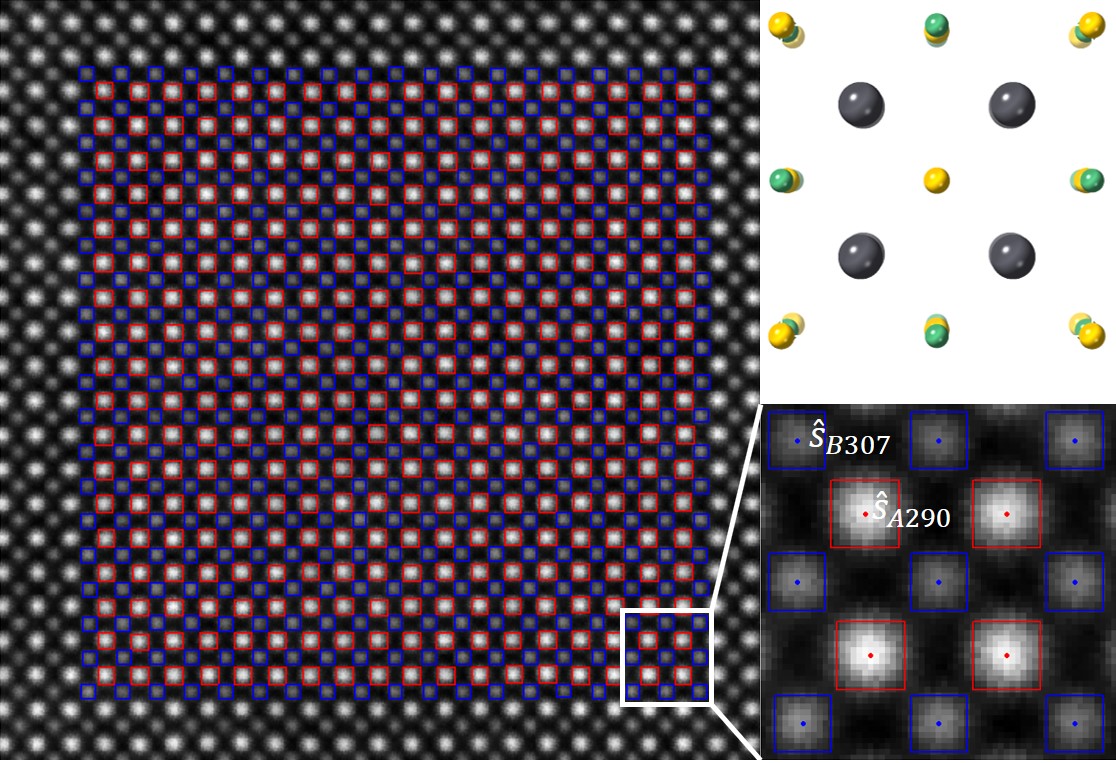}
 \label{fig:STEMimage}
\end{figure}

A common task in the physical sciences is to identify the location and movement of objects of interest via imaging. The locations of these objects often provide information about properties of some system containing the object, so if these location measurements are inaccurate then parameter estimates will be biased. For instance,  astronomers use light intensity at star locations over time to plot light curves and infer rotation periods from these curves \citep{aigrain2016k2sc, douglas2016k2}, or trace orbits of star locations around black holes \citep{schodel2002star}. Another example is estimating a source's contribution of air pollution where the source's location is uncertain , such as \cite{larsen2018impacts}'s study of forest fire emissions on ambient air pollution. Materials scientists study atomic-scale material properties through imaging techniques like scanning transmission electron microscopy (STEM). STEM images of properly aligned crystalline materials show a projection of columns of atoms (Figure \ref{fig:STEMimage}). The locations of these columns are measured with error, which can impact our understanding of material properties. 

From the analysis of atomic resolution STEM images, researchers can determine atom column locations and intensities that reveal a material's local atomic structure and chemical composition, which can govern material properties. Recently, STEM investigations have illustrated how changes in chemical composition of a material modifies its chemical distribution and atomic structure, thereby significantly modifying the material properties \citep{Kumar2019}.  Engineering and controlling material behavior require accurate and precise characterization of chemical and structural relationships \citep{Keen2015}, so it is important that these relationships are properly modeled. In particular, in relaxor ferroelectric materials like the one shown in Figure \ref{fig:STEMimage}, local polarization in the material corresponds to macro-level properties that make the material useful in a variety of applications, including ultrasound imaging, sensors and actuators \citep{Kumar2019}. Polarization is related to displacement of atom columns from their expected position, which in turn may be related to the chemistry of neighboring atom columns. We use a Bayesian framework to model and quantify the uncertainty of the relationship between neighboring chemistry and atom column displacement. 

While the model we develop is in the setting of crystalline materials and STEM microscopy, the underlying techniques could apply to any image containing objects with locations that are measured with error. Our analysis tests hypotheses about the relationship between the positions of neighboring atom columns shown in Figure \ref{fig:STEMimage}. Error from these location measurements can alter this analysis. Therefore, it is important that we account for this measurement error (ME) in our statistical model to make correct conclusions. 

ME in covariates in linear regression settings results in biased parameter estimates that attenuate towards zero \citep{carroll2006measurement}. There are a variety of methods to correct for this bias in models with independent error terms, including regression calibration \citep{carroll1990approximate, gleser1990improvements}, simulation extrapolation (SIMEX) \citep{cook1994simulation}, and Bayesian hierarchical modeling with informative priors on the ME variance based on expert knowledge or repeated measures. \cite{muff2015bayesian} provide a review of Bayesian ME models with several applications and use integrated nested Laplace approximations to carry out their analysis. 

The STEM data in Figure \ref{fig:STEMimage} exhibit spatial dependence, and so we are interested in ME methods for spatial settings. ME methods for spatial statistics have particularly been developed for spatially misaligned data where covariates are observed at locations different from where the response is observed \citep{szpiro2011efficient, gryparis2008measurement}. \cite{li2009spatial} create a spatial linear mixed models ME framework and show that regression coefficient attenuation and variance inflation occur with naive estimates in spatial settings as well. \cite{alexeeff2016spatial} introduce SIMEX for spatial settings where either the data is misaligned or the model is misspecified, and \cite{huque2016spatial} present a spatial analogue to regression calibration. Recently, \cite{tadayon2018spatial} and \cite{tadayon2019non} have developed ME models for non-Gaussian settings by incorporating the ME variance into the spatial covariance. These methods all require knowledge of the ME variance, the ability to estimate it, or assumptions about the ME to make the model identifiable. 

Spatial statistical models incorporate observation locations into the model design via covariates and covariance functions, and thus ME in the locations themselves will impact prediction and inference in these models. There has been some work addressing location ME specifically. Location ME for geostatistics was first explored by \cite{gabrosek2002effect} and \cite{cressie2003spatial}, who developed kriging equations in the context of location ME. \cite{fanshawe2011spatial} developed likelihood-based methods for location ME and \cite{fronterre2018geostatistical} use a composite likelihood approach to speed up these methods and apply them to geomasked data. Again, these methods require knowledge or an estimate of the location ME variance. In imaging applications, we might not have access to information about the ME variance. We develop a model that uses the information in the image itself to infer the variance.

Instead of including informative priors on the ME variance, we expand the model into a hierarchical setting that incorporates every pixel and treats the locations as parameters of the model. The data layer of the hierarchy treats pixel intensities as responses and weights each pixel's contribution to locations of interest by its distance from the location. 
In STEM images, because atom column locations provide information about material chemistry and structure, ME in these locations could bias our understanding of local effects in these materials. Therefore, these images are natural candidates for the described hierarchical framework. Spatial correlation between pixels, however, creates computational issues, as the large size of the image results in an enormous covariance structure and a likelihood that is impossible to compute. Thus, we must approximate the likelihood or the covariance matrix (or both) in order to implement a computationally tractable Bayesian hierarchical model that accounts for ME in the atom column locations.

\cite{heaton2018case} compare the performance of various low rank and sparse covariance/precision approximations for large data sets. Low rank approximations are popular, but \cite{stein2014limitations} showed that contiguous independent block likelihood approximations outperform low rank models when the nugget variance is small and the observations are dense. He points out that the independent contiguous block assumption is troubling, and suggests using composite likelihood methods instead \citep{vecchia1988estimation, varin2011overview, katzfuss2017general, fronterre2018geostatistical}. These STEM images, however, are the ideal candidates for independent blocks. The purpose of using the image is to find the atom column locations and propagate the uncertainty of those locations through our model, and the pixels between the atom columns do not contain information about the center of those columns. We see in Figure \ref{fig:STEMimage} that atom columns appear as bright circles in the images with dark, low-information areas around them. We put boxes around the atom columns and treat the observations in one box as independent from the observations in another. We discard the observations outside of the boxes since they contain little information about atom column positions. Thus, we have a collection of non-contiguous boxes that we can reasonably assume are independent, as opposed to the contiguous blocks described by \cite{stein2014limitations}. 

This approach differs from other methods because it uses the data in the image to account for the ME, instead of estimating it or assuming something about the underlying process. Additionally, the computational time scales linearly with the number of atoms, making it feasible to use for very large data sets. Furthermore, while \cite{den2005maximum} and \cite{van2005maximum} characterize structural parameters in atom columns using frequentist methods, they treat residuals as uncorrelated. We incorporate spatial correlation between pixels and atom columns into our model, use Bayesian methods to quantify uncertainty in our parameters, and take advantage of a hierarchical framework to perform inference on parameters that characterize physical and chemical relationships between atoms columns. 

The rest of this article proceeds as follows. In Section 2, we explain how we collected the data, how the intensity of the atom columns relates to the chemistry of those columns, and introduce our notation. We describe the hierarchical model and approximate likelihood of the data layer in Section 3. In Section 4 we discuss the Markov chain Monte Carlo (MCMC) setup. We compare the hierarchical model with standard spatial and simple linear regression models in Section 5 via a simulation study. We apply and compare these methods on collected STEM image data in Section 6, finding a negative relationship between atom column displacement and the weighted intensity of their neighbors. We conclude in Section 7.

\section{STEM Imaging Data and Description}
Pb(Mg$_{1/3}$Nb$_{2/3}$)O$_3$ (PMN) is a relaxor ferroelectric material with perovskite structure. Perovskite crystals have two main types of atom sites, generically called A- and B-sites. In PMN, the A-sites are exclusively lead, while one third of the B-sites contain magnesium and two thirds contain niobium on average. High angle annular dark-field (HAADF) STEM images allows us to directly view and identify columns of A- and B-sites based on intensity. The intensity of a pixel is a unitless representation of the flux of electrons that hit the detector at the pixel.  The intensity is dependent on the atomic number (Z) and the thickness of a sample. Assuming a uniformly thick specimen, an atom column consisting of Pb (Z = 82) will appear brighter than a column containing Mg (Z = 12) and Nb (Z = 41) \citep{lebeau2008experimental}. B-site pixel intensities increase with the proportion of the column that is Nb, as it has a higher atomic number than Mg.

Figure \ref{fig:STEMimage} shows a $551\times 551$ pixel image with $19^2=361$ identified B-site columns (blue boxes) and $18^2=324$ identified A-site columns (red boxes). The Appendix describes the atom column identification and location estimation processes. Atomic arrangement in relaxor ferroelectrics such as PMN drive their unique material properties. Relaxor ferroelectrics and their properties are highly sensitive to their chemical make up as evidenced by a recent study that demonstrated a material property of interest could be doubled by substituting $<$1\% of one constituent element for another \citep{Li2019}. Understanding how individual atoms influence their surrounding structure is important for understanding the origin of material properties, and in turn, how to engineer them for even greater properties \citep{Keen2015}. 

\begin{figure}
 \centering
  \caption{Diagram of negative A-site displacement in response to the difference in intensity-weighted and unweighted averages of neighboring B-sites. $\mathbf{s}_{Bj}$ are the B-sites neighboring A-site $\mathbf{s}_{A1}$, $\mathbf{w}_{A1}$ is the intensity-weighted average of the locations of the B-sites and $\mathbf{u}_{A1}$ is the unweighted average. The black B-site is more intense than the three gray B-sites.}
 \includegraphics[page=1,width=.75\textwidth]{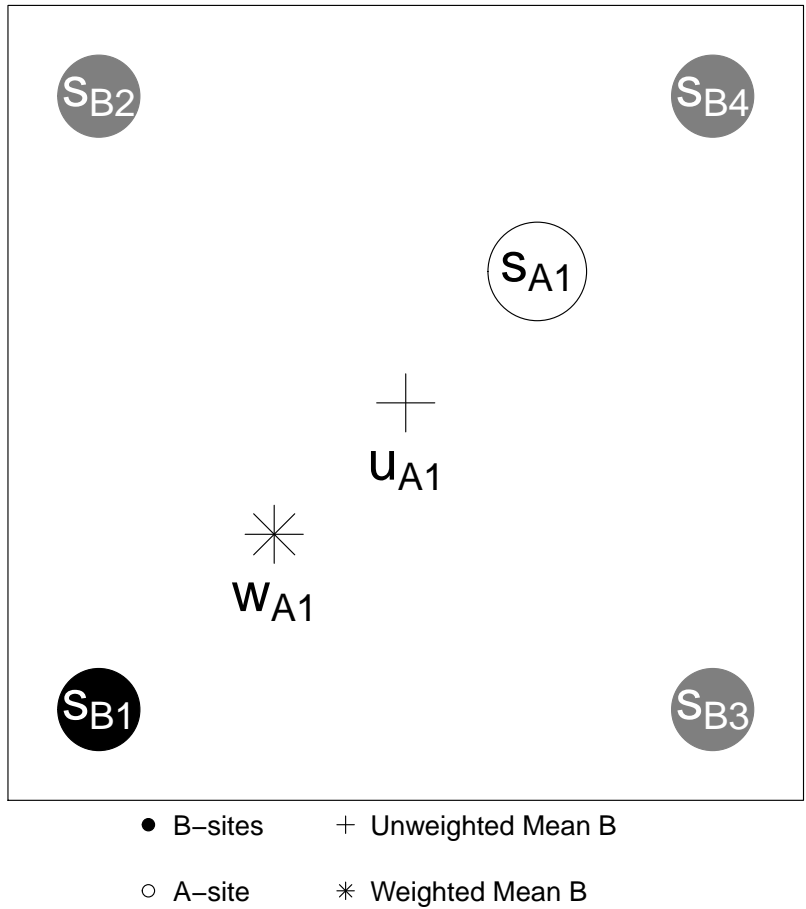}
 \label{fig:COM}
\end{figure}

We are particularly interested in the relationship between the intensity of the B-site columns and the displacement of the neighboring A-site column from its expected location. We introduce the notation and framework for modeling this relationship in Figure \ref{fig:COM}. We denote the $j^{th}$ B-site column location and the $k^{th}$ A-site column location as $\mathbf{s}_{Bj}$ and $\mathbf{s}_{Ak}$, respectively. In Figure $\ref{fig:COM}$, $\mathbf{s}_{B1}, \dots, \mathbf{s}_{B4}$ are the locations of the B-site columns that are the neighbors of the A-site column at $\mathbf{s}_{A1}$. Here, the column at $\mathbf{s}_{B1}$ has a higher intensity than the other three B-site columns. According to the perovskite crystal structure of PMN, the location of the A-site column should be at the unweighted mean location of the neighboring B sites, denoted as $\bf{u}_{A1}$ in the figure. We model the displacement of the A-site, $\mathbf{s}_{A1}-\mathbf{u}_{A1}$  as a function of $\mathbf{w}_{A1}-\mathbf{u}_{A1}$, where $\mathbf{w}_{A1}$ is the intensity-weighted average of the neighboring B-site locations. In Figure \ref{fig:COM} there is a negative relationship between displacement and $\mathbf{w}_{A1}-\mathbf{u}_{A1}$.

The magnitude of the effect of this relationship may be small, possibly less than one pixel. Therefore, even small errors in estimates of atom column locations due to the algorithm or to the resolution of the image may result in attenuation of the estimate of this effect. Thus, we develop a hierarchical model to account for these errors. In the hierarchical model, the estimated locations are used as initial values and as references to form the non-contiguous blocks in the approximate likelihood, but are not treated as the true locations.

%\subsection{Finding the Atom Column Locations}
%\subsection{Data Reduction and Likelihood Approximation}
%We would ideally like to use the intensity stored with each pixel so that all of the information in the image is in our model, but this is not computationally feasible. These images are large, sometimes $1024\times 1024 \approx 1,000,000$ pixels. We are interested in spatially correlated error between these pixels, but a $1,000,000 \times 1,000,000$ covariance matrix is difficult to store and calculating determinants and inverses to evaluate the likelihood is impossible. 

%In reality, however, not all pixels provide the same amount of information about the atom columns. There are many pixels that are nearly dark, and we can clearly see that there is space between the atom columns. We therefore approximate the likelihood by putting boxes around the atom columns and treating pixels in separate boxes as independent. 

%Figure 1 shows the boxes around the A-sites in red and around the B-sites in blue. We place the boxes based on the calculated initial atom column locations. Since the A-sites are larger than the B-sites as a function of the size of their corresponding atoms, the A-site boxes have a width of $11$ pixels and the B-site boxes have a width of $9$ pixels. Now we have two manageable covariance matrices, one of size $121 \times 121$ and the other size $81 \times 81$. We only need to calculate the distance matrices for these once since they will be the same for each box of the same size (find reference for blocks).
\section{Model Description}
We use the Bayesian hierarchical framework for our statistical model. The data layer encapsulates the relationship between the intensities associated with each pixel and the atom column locations and intensities. The process layer models the association between the displacement of the A-site locations from their expected position and the intensities of neighboring B-sites. We compare the hierarchical model to the described spatial and simple linear regression models with fixed atom column locations. In this section, we explain the model in detail, but also provide a reference of the names, descriptions and priors of the parameters in Table $\ref{table:parameters}$.

\begin{table}
\begin{center}
\caption{Description of parameters, hyperparameters and associated prior distributions for the hierarchical model. The hyperparameters for the variance of the $\beta_{ij}$ come from setting the mean to be the sample variance of the OLS estimates and the variance to be $25^2$, with $a_i = \frac{\hat{\sigma}^2_{\beta_i}}{25^2} + 2$ and $b_i =\hat{\sigma}^2_{\beta_i}(\frac{\hat{\sigma}^2_{\beta_i}}{25^2} + 1)$. $\mu_{Aj}$ is the mean defined in equation \ref{equation:s_A} and and $\overset{\sim}{\mu}_{Bj}$ is the grid location described in Section 3.1.3.}
\begin{tabular}{p{2.5cm}p{5.5cm}p{4cm}}
Parameter(s) & Description & Prior \\
\hline
$\beta_0$ & Intercept for pixel intensity & $\text{Normal}(0,1000^2)$\\
$\beta_{ij}$ & Slope associated with pixel intensity for atom $j$ of type $i$ & $\text{Normal}(\mu_{\beta_i}, \sigma^2_\beta)$ \\
$\mu_{\beta_A}, \mu_{\beta_B}$  & Hyperparameters; means of the A- and B-site $\beta$'s & $\text{Normal}(0,1000^2)$\\
$\sigma^2_{\beta_A}, \sigma^2_{\beta_B}$ & Hyperparameter; variance of $\beta$'s & $\text{InvGamma}(a_i, b_i)$\\
$\psi_A, \psi_B$ & Bandwidth for A- and B- site intensities & $\text{LogNormal}(0,100)$ \\
$\sigma^2$ & Pixel intensity variance & $\text{InvGamma}(.01,.01)$\\
$r, r_{\text{pix}}$ & Proportion of variance that is spatial for atoms and pixels, respectively& $\text{Uniform}(0,1)$\\
$\rho, \rho_{\text{pix}}$ & Spatial range for atoms and pixels, respectively & $\text{LogNormal}(0,100)$\\
$\mathbf{s}_{Aj}$ & Coordinates of the $j^{th}$ A-site & $\text{Normal}(\mu_{Aj}, \sigma^2_A\mathbf{I}_2)$\\
$\mathbf{s}_{Bj}$ & Coordinates of the $j^{th}$ B-site & $\text{Normal}(\overset{\sim}{\mathbf{\mu}}_{Bj}, \sigma^2_B\mathbf{I}_2)$\\
$\alpha_0, \alpha_1$ & Intercept and slope for A-site displacement vs difference of weighted and unweighted B-site averages & $\text{Normal}(0, 1000^2)$\\ $\sigma^2_A, \sigma^2_B$ & A-site and B-site variance & $\text{InvGamma}(.01,.01)$
\end{tabular}
\label{table:parameters}
\end{center}
\end{table}
\subsection{Data Layer}
In the data layer we model the relationship between image intensity $Y(\mathbf{p})$ at the $2\times 1$ pixel location vector $\mathbf{p}$ and the locations of the atom columns. Let $\mathbf{s}_{ij}$ be the $2\times 1$ coordinate vector of the $j^{th}$ atom column of type $i \in \{A, B\}$ and $\beta_{ij}$ be the corresponding intensity parameter for that atom column. Let $\beta_0$ represent the background intensity, $\psi_i$ be the bandwidth parameter for type $i$ atoms and $\epsilon(\mathbf{p})$ be a spatial error term. 

The model for the observed intensities is
\begin{equation}
 Y(\mathbf{p}) = \beta_0 + \sum_{i \in \{A,B\}}\sum_{j=1}^{N_i}\beta_{ij}\exp\Big(-\frac{\lVert \mathbf{p} - \mathbf{s}_{ij} \rVert^2}{2\psi_i^2}\Big) + \epsilon(\mathbf{p}),
 \label{equation:data_layer}
\end{equation}
where $\lVert \cdot \rVert$ is the Euclidean norm. The expected intensity decays from the atom column location following a Gaussian kernel to place higher intensity value on pixels closer to the center of the nearest atom column. We justify the use of a Gaussian kernel by fitting Gaussian curves to the mean of the horizontal and vertical traces through the atom columns in Figure \ref{fig:STEMimage}. The resulting fits align with the mean traces in both directions, strongly suggesting that a Gaussian kernel is appropriate. We provide a plot and further discussion of these findings in the Supplementary Information.

For the residuals, let $\sigma^2$ be the variance, $r_{\text{pix}}$ be the proportion of variance that is spatial, and $\rho_{\text{pix}}$ be the spatial range. The residuals $\epsilon(\mathbf{p})$ follow a Gaussian process denoted $GP(\sigma^2, r_\text{{pix}},\rho_{\text{pix}})$ with mean $0$ and exponential covariance function
\begin{equation}
C\big(\epsilon(\mathbf{p}), \epsilon(\mathbf{p}')\big) = \sigma^2\Big[(1-r_{\text{pix}})I(\mathbf{p}=\mathbf{p}')+r_{\text{pix}}\exp\big(-\frac{\lVert \mathbf{p} - \mathbf{p}'\rVert}{\rho_{\text{pix}}}\big)\Big],
\label{equation:cov}
\end{equation}
where $I(\cdot)$ is the indicator function. The exponential covariance function is a part of the desirable Mat\'ern class of covariance functions where the smoothness parameter is $\frac{1}{2}$ \citep{gelfand2010handbook}. We use exponential covariance functions in both the data and process layers after examining  empirical variograms of the residuals of OLS estimates of the data layer model and fitting exponential covariance functions to the variograms. We provide the details of this justification in the Supplementary Information.

The model in (\ref{equation:data_layer}) is not feasible from a computational perspective, so here we present a justifiable approximation. We approximate our model as independent across windows surrounding the atoms, as shown in Figure \ref{fig:STEMimage}. We only consider pixels within square windows $\mathbf{W}_{ij}$ around column $\mathbf{s}_{ij}$, thus moving from the contiguous blocks described by \cite{stein2014limitations} to non-contiguous blocks of equal sizes for each atom column type. Since the atom columns outside of the window are far from the pixels inside the window, we treat their contributions as negligible. This is justified for multiple reasons. First, the bandwidths for the Gaussian kernels are narrow enough that nearby atom columns will only minimally contribute to the intensities of the pixels near the atom column centers. Second, the spatial range in the empirical variograms is small (see Supplementary Information). Third,	from an error structure perspective, we are pooling the information across sites to find the correlation parameters, so while each individual window might not be enough to cover these, the combination can with low bias. Finally, the error from approximating via independent blocks is spatial, so this error will be absorbed into the spatial error term when fitting the model.

Let $Y(\mathbf{p}_{ijk})$ be the intensity of the $k^{th}$ pixel in window $\mathbf{W}_{ij}$ and $\mathbf{p}_{ijk} \in \mathbf{W}_{ij}$ be the $2\times 1$ location vector of that pixel. Then, we approximate our model from (\ref{equation:data_layer}) as 
\begin{equation}
 Y(\mathbf{p}_{ijk}) = \beta_0 + \beta_{ij}\exp\Big(-\frac{\lVert \mathbf{p}_{ijk} - \mathbf{s}_{ij} \rVert^2}{2\psi_i^2}\Big) + \epsilon(\mathbf{p}_{ijk}),
    \label{equation:appx_datalayer}
\end{equation}
The covariance for pixels within $\mathbf{W}_{ij}$ follows (\ref{equation:cov}), and is $0$ for pixels that are not in the same window.
\subsection{Process Layer}
The objective of our study is to test whether the displacement of the A-sites from the unweighted center of the neighboring B-sites relates to the intensity of the B-sites. For the $B$ neighboring B-sites of the $j^{th}$ A-site, the unweighted center is 
\begin{equation}
\mathbf{u}_{Aj}=\frac{1}{B}\sum_{k\sim j}\mathbf{s}_{Bk}. 
\label{equation:unweighted}
\end{equation}
where $k\sim j$ denotes the $k^{th}$ neighbor of the $j^{th}$ site. The $\beta_{Bk}$ parameters in the data layer are the intensities of the B-sites, so the weighted center is
\begin{equation}
\mathbf{w}_{Aj} = \frac{\sum_{j\sim k}\beta_{Bk}\mathbf{s}_{Bk}}{\sum_{k\sim j}\beta_{Bk}}.
\label{equation:weighted}
\end{equation}

For $l\in \{x,y\}$, let $s_{Ajl}, u_{Ajl}$, and $w_{Ajl}$ be the $l^{th}$ coordinates of $\mathbf{s}_{Aj}, \mathbf{u}_{Aj}$, and $\mathbf{w}_{Aj}$, respectively. The process layer models the A-site column locations, conditioned on all B-site column locations $\mathbf{s}_B = \{\mathbf{s}_{Bk} \text{ for all } k\}$:
\begin{equation}
s_{Ajl}|\mathbf{s}_{B}, \boldsymbol{\beta}, \alpha_0, \alpha_1, \sigma^2_A = u_{Ajl} + \alpha_0 + \alpha_1(w_{Ajl} - u_{Ajl}) + \varepsilon(s_{Ajl}).
\label{equation:s_A}
\end{equation}
The residuals $\varepsilon$ are independent between x- and y- coordinates and follow a mean-zero Gaussian process $GP(\sigma^2_A, r, \rho)$ with the exponential covariance structure defined in (\ref{equation:cov}), where $\sigma^2_A$ is the A-site variance $r$ is the proportion of variance that is spatial and $\rho$ is the spatial range.

The $2\times 1$ vector $\mathbf{s}_{Aj}-\mathbf{u}_{Aj}$ is the $x-$ and $y-$ displacement of the A-site from the central position, and the displacement resembles simple linear regression with covariate $\mathbf{w}_{Aj} - \mathbf{u}_{Aj}$. The intercept parameter is $\alpha_0$. The slope parameter $\alpha_1$ models the linear relationship between displacement of the A-site and the difference between the weighted and unweighted averages of its neighboring B-sites. In other words, a relatively high-intensity B-site is associated with greater A-site displacement.  This model frames the study's objective as a test of whether $\alpha_1 = 0$.  

We model the B-site locations as 

\begin{equation}
 \mathbf{s}_{Bj}|\mathbf{\tilde s}_{Bj}, \sigma^2_B \overset{\text{ind.}}{\sim} \mathbf{N}(\mathbf{\tilde s}_{Bj}, \sigma^2_B\mathbf{I}_2),
 \label{equation:s_B}
\end{equation}
where $\mathbf{\tilde s}_{Bj}$ is the expected location of the B-site based on the symmetric properties of the crystal structure of the material. $\sigma^2_B$ controls B-site displacement from the crystal structure. We treat the B-sites as uncorrelated because we expect the deviation of the sites from their expected location on the crystal lattice to be small. The knowledge of the crystal structure grounds our model around where the B-sites should be and is propagated through (\ref{equation:s_A}) via the unweighted and weighted averages of the B-sites in the covariates and the A-site displacement. This prior structural knowledge ensures the model is identifiable.
\subsection{Prior Layer}
In general, we choose weakly informative priors for our parameters. %We do make use of OLS estimates for the $\beta$'s along with a hyperparameter for the variance of the $\beta$'s, and use knowledge of the crystal structure of the material in the calculation of the prior mean of the B-sites. 
The means for the B-sites are on an equispaced grid $\overset{\sim}{\mathbf{\mu}}_{B}$ calculated from the corner sites, which corresponds to the perovskite structure of PMN. We use OLS estimates of the $\beta_{ij}$ to ground the hyperparameters $\sigma^2_{\beta_i}$ at reasonable values. In particular, we set the mean for $\sigma^2_{\beta_i}$ at the sample standard deviation of the OLS estimates of $\beta_{ij}$ and the variance of $\sigma^2_{\beta_i}$ at $25^2$. We let the priors for $\sigma^2_i$ follow Inverse Gamma distributions, so we solve for the shape and rate parameters based on the mean and variance settings. 

\section{Computing}
We use Gibbs and Metropolis sampling in an MCMC framework to sample from the joint posterior distribution of the parameters. The description of the prior distributions of the hierarchical model parameters is in Table \ref{table:parameters}. For the non-hierarchical models, the regression coefficients $\alpha_0$ and $\alpha_1$ have conjugate $N(0,1000^2)$ priors. We also use a Gibbs sampler for variance $\sigma^2_A$, with an InverseGamma$(0.01, 0.01)$ conjugate prior. In the spatial linear regression model, we use Metropolis samplers for the correlation parameters $r$ and $\rho$ with Uniform$(0,1)$ and LogNormal$(0,10)$ priors, respectively. 

The hierarchical model contains $3(N_A+N_B) + 16$ parameters, where $N_i$ is the number of type-$i$ columns. As such, the number of parameters scale linearly with the number of atom columns. To mitigate the large computational burden we break the image into independent blocks, placing boxes around each column as described in Section 2. The boxes must not overlap, or we will count pixels more than once in our analysis and have an invalid model. Therefore, the size of the boxes is important, as they must contain the atom column while not overlapping with the other boxes. It is helpful to orient the image so that it is not necessary to rotate the boxes to be in line with the rows of atom columns. 

After selecting box half-widths of $h_A$ and $h_B$ for the A- and B-sites, respectively, we create the boxes by rounding the estimates for the atom column locations to the nearest pixel, then adding and subtracting the half-widths from the x- and y-coordinates to get the pixels inside of the box. Thus we have square boxes of width $2h_i + 1$ around each atom column of interest, where $i \in \{A,B\}$. The approximate likelihood is then

\begin{equation}
 p(\mathbf{Y}|\mathbf{\Theta}) \approx \prod_{i\in \{A,B\}}\prod_{j=1}^{N_i}p^*(\mathbf{Y}_{ij}|\mathbf{s}_{ij},\mathbf{\Theta}),
 \label{equation:approx_like}
\end{equation}
where $p^*(\cdot)$ is the density from the approximate model in (\ref{equation:appx_datalayer}), $\mathbf{Y}_{ij}$ is the vector of pixels in window $\mathbf{W}_{ij}$, and $\Theta$ is the vector of parameters other than the location of the $j^{th}$ atom column of type $i$. 

Because these boxes are the same size for each atom type, we need only to compute the two pixel-pixel distance matrices (one for A-sites and one for B-sites) for the covariance matrices in the likelihood, making likelihood calculations very efficient. See the Supplementary Materials for the derivations of the sampler updating steps. The code for our MCMC algorithm, simulations, and figures is available at https://github.com/reich-group/HierarchicalSTEM.  

\section{Simulation Study}
We simulate 100 data sets for each model setting, drawing 10,000 posterior samples for each data set after a 10,000 iteration burn-in period. We compare the hierarchical model against the spatial and simple linear regression models with fixed atom column locations described in Section 5.2. The window half-widths are 6 pixels for the A-sites and 5 pixels for the B-sites. 
\subsection{Data Generation}
We generate data to have similar properties to the real data plotted in Figure \ref{fig:STEMimage}. We also consider simulations with slightly different true parameters to understand the operating characteristics of the proposed method.
\subsubsection{Atom Column Locations}
We first draw $19^2$ B-sites from a normal distribution where the mean is a grid of points $40$ pixels apart and the standard deviation $\sigma_B = 0.25$. To simulate the locations of the corresponding $18^2$ A-sites, we first need to generate the $\beta$'s. We set $\beta_0 = 87$, and independently draw $\beta_{ij} \sim N(\mu_{\beta_i}, \sigma^2_{\beta_i})$, where $\mu_{\beta_A} = 3060$, $\mu_{\beta_B} = 1425,$ and $\sigma^2_{\beta_A}=\sigma^2_{\beta_B} = 150$. Letting the A-site distance matrix $d$ be defined by the unweighted means of neighboring B-sites in the mean grid, with $\alpha_0 = -0.08$, $\alpha_1 = -0.15$ %for setting 1 and $-0.5$ for setting 2
, $\sigma_A = 0.4, r = 0.73,$ and $\rho = 100$, we draw the A-sites from the distribution defined in (\ref{equation:s_A}).

\subsubsection{Pixel Intensities}
We examine five model settings by fixing correlation parameter $r_{pix} = 0.57$ and varying intensity standard deviation $\sigma$ between 140, 220, and 300 for the first three settings. For the last two, we fix $\sigma = 140$ and change $r_{pix}$ to be 0.7 and 0.9. We set the bandwidth parameters $\psi_A = 4.3$ and $\psi_B = 3.7$ and pixel spatial range $\rho_{pix} = 5.5$. We draw the pixel intensity values based on (\ref{equation:data_layer}) for pixels within a $2(h_i+2)+1$ width box around each atom of type $i$. The purpose of this is to ensure that the boxes with half-widths $h_i$ drawn around the estimated atom locations contain pixels that follow the proper distribution. The remaining pixel intensities come from an i.i.d $N(\beta_0, 25)$ distribution. 

\subsubsection{Initial Atom Column Locations}
The algorithm described in the Appendix chooses the initial atom column locations by first finding the intensity-weighted average of the nearby pixels and then using nonlinear least squares to refine this estimate. Because we already know the general location of each atom based on the boxes, we skip the Normalized Cross-Correlation (NCC) step, using the pixels inside each corresponding box. In most cases, the nonlinear least squares fit and the intensity-weighted average produce the same location. 

\subsection{Non-Hierarchical Models}
Bayesian spatial and simple linear regression using fixed atom column locations provide faster and more straightforward analyses at the cost of bias and variance inflation from naive parameter estimates. We estimate the atom locations using the non-linear least squares method described in the Appendix, and assume them to be known for the remainder of the analysis. We modify the models from \cite{CabralThesis} by combining the x- and y-displacements into one vector. The new models are of the form
\begin{equation}
 \mathbf{\delta}(\mathbf{s}_{Aj}) = \alpha_0 + \alpha_1\mathbf{\Psi}(\mathbf{s}_{Aj}) + \mathbf{\epsilon}(\mathbf{s}_{Aj}),
 \label{equation:simple}
\end{equation}
where $\mathbf{\delta}(\mathbf{s}_{Aj}) = \mathbf{s}_{Aj} - \mathbf{u}_{Aj}$, $\mathbf{\Psi}(\mathbf{s}_{Aj}) = \mathbf{w}^*_{Aj} - \mathbf{u}_{Aj}$, and $\text{Cov}(\mathbf{\epsilon}(\mathbf{s}_{Aj})) = \sigma^2_A\mathbf{I}_2$. $\mathbf{u}_{Aj}$ is defined in (\ref{equation:unweighted}) and 
\begin{equation}
 \mathbf{w}^*_{Aj} = \frac{\sum_{j\sim k}\hat{Y}(\mathbf{\hat{s}_{Bk}})\mathbf{\hat{s}}_{Bk}}{\sum_{j\sim k}\hat{Y}(\mathbf{\hat{s}}_{Bk})} ,
\label{weighted_simp}
\end{equation}
which is the analogue for (\ref{equation:weighted}) when every pixel is not in the model. $\hat{Y}(\mathbf{\hat{s}}_{Bk})$ is the intensity found from the nonlinear least squares fit in the Appendix. The covariance structure for the spatial linear regression model is the same as for the hierarchical model and the residuals for the simple linear regression model are i.i.d. normal with mean 0 and variance $\sigma^2_A$. 

\subsection{Results}
We are primarily interested in the slope parameter $\alpha_1$. Table \ref{table:sim_alpha} displays the bias of the posterior means, mean posterior standard deviation, coverage and estimated Mean Squared Error ($\widehat{MSE}$) for $\alpha_1$ in all model settings. The hierarchical model has the highest coverage and lowest $\widehat{MSE}$ for $\alpha_1$ compared to simple and spatial linear regression for every setting. The hierarchical model captures the true regression coefficient, while the posterior mean estimator of $\alpha_1$ in the spatial and simple linear regression models attenuates towards zero, as expected from the ME literature. The attenuation contributes to poor coverage in the naive models, whereas the hierarchical models perform well until the intensity standard deviation $\sigma$ increases drastically. We see the hierarchical model performance decline slightly when $\sigma=220$, and perform much worse when $\sigma=300$. We also examined the sensitivity of the model to the choice of $a_i$ and $b_i$, the hyperparameters on the variance $\sigma^2_{\beta_i}$ of the intensity parameters $\beta_{ij}$, where $i\in \{A, B\}$. These parameters depend on OLS estimates of the intensity parameters as well as a chosen variance, and we found model performance to be robust to increases and decreases of 50\% in this variance. We also found that coverage of $\alpha_1$ slightly dropped when we decreased the window size from $13^2$ pixels to $11^2$ pixels for A-sites and from $11^2$ pixels to $9^2$ pixels for B-sites (see Supplementary Information).

Table \ref{table:simresults} displays the results of more parameters from the initial model setting. For the parameters common between the three models, the hierarchical model has the best coverage, though the spatial linear regression model has tighter posteriors for the correlation parameters, which is reflected in MSE estimates. The data layer parameters have less than 95\% coverage, but the bias and means of the posterior standard deviations show that they are close to the truth for the most part. The low coverage may be explained by the pixels inside the windows not capturing all of the information in the model. However, the parameters of interest are in the process layer, not the data layer, and this model sees better performance in the process layer parameters than the spatial and simple linear regression models.

\begin{table}[]
\caption{Summary of simulation study performance for estimating $\alpha_1 = -0.15$ under various parameter settings for simple linear regression (SimpLR), spatial linear regression (SpatLR), and our new Bayesian hierarchical model (Hierarch). We simulated 100 data sets for each parameter setting. Monte Carlo standard errors are in parentheses.}

\resizebox{\textwidth}{!}{
\begin{tabular}{lllllll}
Statistics     & Model & $r_{pix}=0.57$  & $r_{pix} = 0.57$, & $r_{pix} = 0.57$, & $r_{pix} = 0.7$, & $r_{pix} = 0.9$\\
 & & $\sigma = 140$ & $\sigma = 220$ & $\sigma = 300$ & $\sigma = 140$ & $\sigma = 140$\\
\hline
    & SimpLR & 0.037 (0.0013) & 0.070 (0.0013)      & 0.092 (0.0013)      & 0.042 (0.0013)      & 0.052 (0.0013)      \\
Bias & SpatLR & 0.037 (0.0013) & 0.069 (0.0013)      & 0.091 (0.0013)     & 0.042  (0.0012)     & 0.052  (0.0012)     \\
       & Hierarch & -0.002 (0.0016) & 0.001 (0.0020)     & 0.046 (0.0020)      & -0.004 (0.0018)      & -0.004 (0.0020)      \\
\hline
     & SimpLR & 0.016 (0.0001) & 0.015 (0.0001)     & 0.015 (0.0001)      & 0.016 (0.0001)      & 0.015 (0.0001)      \\
 Mean Post. SD & SpatLR & 0.012 (0.0001) & 0.012 (0.0001)      & 0.012 (0.0001)      & 0.012 (0.0001)      & 0.012 (0.0001)      \\
       & Hierarch & 0.017 (0.0002) & 0.023  (0.0003)     & 0.025 (0.0003)      & 0.018 (0.0002)      & 0.020 (0.0002)     \\
\hline
    & SimpLR & 37 (4.8)      & 0 (0) & 0 (0)      & 21 (4.1)      & 3 (1.7)       \\
 \% Coverage & SpatLR & 22 (4.1)      & 0 (0) & 0 (0)     & 7 (2.7)      & 0 (0)       \\
       & Hierarch & 95 (2.2)       & 94 (2.4) & 51 (5.0)       & 97 (2.0)      & 97 (1.7)      \\
\hline
 & SimpLR & 0.15 (0.011)  &0.50 (0.018)     & 0.86 (0.024)      & 0.20 (0.011)      & 0.29 (0.014)     \\
 $\widehat{MSE}\times 100$ & SpatLR & 0.15 (0.010)      &0.49 (0.018) & 0.85 (0.023)      & 0.19 (0.010)     & 0.28 (0.013)      \\
       & Hierarch & 0.03 (0.004)     &0.05 (0.006) & 0.25 (0.019)      & 0.03 (0.005)     & 0.04 (0.005)      
\end{tabular}}
\label{table:sim_alpha}
\end{table}

\begin{table}[]
\begin{center}
\caption{Simulation study results for 100 simulated data sets with $\alpha_1 = -0.15$ for simple linear regression (SimpLR), spatial linear regression (SpatLR), and our new Bayesian hierarchical model (Hierarch). Coverage is the percent of $95\%$ highest posterior density credible intervals that contain the parameter.}
\resizebox{\textwidth}{!}{
\begin{tabular}{c|rrrrrrrrrr}
   &    &  &  & & Mean & &    & &\\
   &    &  &  & & Post. & & Coverage    & &  &\\
Parameter & Model  & Truth & Bias & (SE)& SD & (SE) &(\%) & (SE)& $\widehat{MSE}$ & (SE)\\
\hline
$\alpha_0$ & SimpLR  & -0.08 & 0.019 & (0.008) & 0.018 & (0.0001) & 40 & (4.9)& 0.0064 & (0.0001)\\
  & SpatLR  & & -0.018 & (0.007) & 0.070 & (0.0017) & 92 & (2.7) & 0.0055 & (0.0008)\\
  %& Hierarch. Small & & -0.05 & & 0.092 & & 94.6  & 0.078 \\
  & Hierarch & & -0.034 & (0.007) & 0.096 & (0.0046) & 95 & (2.2) & 0.0061 & (0.009)\\
\hline
%$\alpha_1$ & SimpLR  & -0.15 & -0.11 & (0.001) & 0.016 &  & 38.4  & 0.039 \\
 %  & SpatLR  & & -0.11 & (0.001) & 0.012 & & 23.2  & 0.038 \\
  %& Hierarch. Small & & -0.14 & & 0.016 & & 88.4  & 0.019 \\
 % & Hierarch. Large & & -0.15 & (0.002) & 0.017 & & 95.5  & 0.016 \\
$\sigma_A$ & SimpLR  & 0.4 & 0.053 & (0.002) & 0.013 & (0.0001) & 6 & (2.4)& 0.0032 & (0.0002)\\
  & SpatLR  & & 0.064 & (0.002) & 0.027 & (0.0001) & 13 &(3.4) & 0.0045 & (0.0003)\\
  %& Hierarch. Small & & 0.42 & & 0.046 & & 92.0  & 0.055 \\
  & Hierarch & & 0.034 & (0.005) & 0.049 & (0.0002) & 94 &(2.4) & 0.0034 & (0.0008)\\
\hline
 %$r$ & SimpLR  & 0.73 & $-$ & $-$ & $-$ & $-$ & $-$  & $-$ \\
 $r$ & SpatLR  & 0.73 & -0.048 & (0.008) & 0.082 & (0.0015) & 88 & (3.2) & 0.0091 & (0.0016)\\
  %& Hierarch. Small & & 0.62 & & 0.086 & & 84.8  & 0.130 \\
  & Hierarch & & -0.114 & (0.007) & 0.085 & (0.0015) & 80 & (4.0) & 0.0174 & (0.0020)\\
\hline
%$\rho$ & SimpLR  & 100 & $-$ & $-$ & $-$ & $-$ & $-$  & $-$ \\
 $\rho$  & SpatLR  & 100 & -5.5 & (3.2) & 31.0 & (1.30) & 86 & (3.5)  & 1053 & (19.3)\\
  %& Hierarch. Small & & 155 & & 91.6 & & 97.3  & 110 \\
  & Hierarch & & 64.5 & (9.9) & 98.4 & (10.11) & 97 & (1.7) & 1385 & (386.4)\\
\hline
$\beta_0$ & Hierarch & 87 & 7.59 & (0.45) & 4.78 & (0.014) & 67 & (4.7) & 77.4 & (7.67)\\

$\beta_{A100}$ & Hierarch & 3006.21 & 63.04 & (13.6) & 79.4& (0.017) & 69 & (4.6) & 22204 & (3571)\\

$\sigma$ & Hierarch & 140 & -2.22 & (0.13) & 1.18& (0.001) & 47 &(5.0) & 6.62 & (0.675)\\

$\psi_A$ & Hierarch & 4.3 & -0.01 & (0.0007) & 0.008& (0.0000) & 79 &(4.1) & 0.0001 & (0.00002)\\

$r_{pix}$ & Hierarch & 0.57 & -0.01 & (0.0008) & 0.008& (0.0000) & 48 &(5.0) & 0.0003 & (0.00003)\\

$\rho_{pix}$ & Hierarch & 5.5 & -0.33 & (0.024) & 0.217& (0.0021) & 63 &(4.8) & 0.17 & (0.019)\\

\end{tabular}}
\label{table:simresults}
\end{center}
\end{table}

\section{STEM Image Analysis}
The PMN image in Figure \ref{fig:STEMimage} contains $19^2$ A-sites and $18^2$ B-sites for analysis. We run the MCMC for each model for 90,000 iterations after a 10,000 iteration burn-in and check convergence visually via trace plots. We compare the hierarchical model with half-width 6 for the A-sites and 5 for the B-sites against the spatial and simple linear regression models described in Section 5.2.

\begin{figure}
 \centering
  \caption{Posterior densities for the five common parameters between the hierarchical, spatial linear regression, and simple linear regression models. The regression parameter $\alpha_1$ attenuates towards zero in the simple and spatial linear regression models.}
 \includegraphics[page=1,width=1\textwidth]{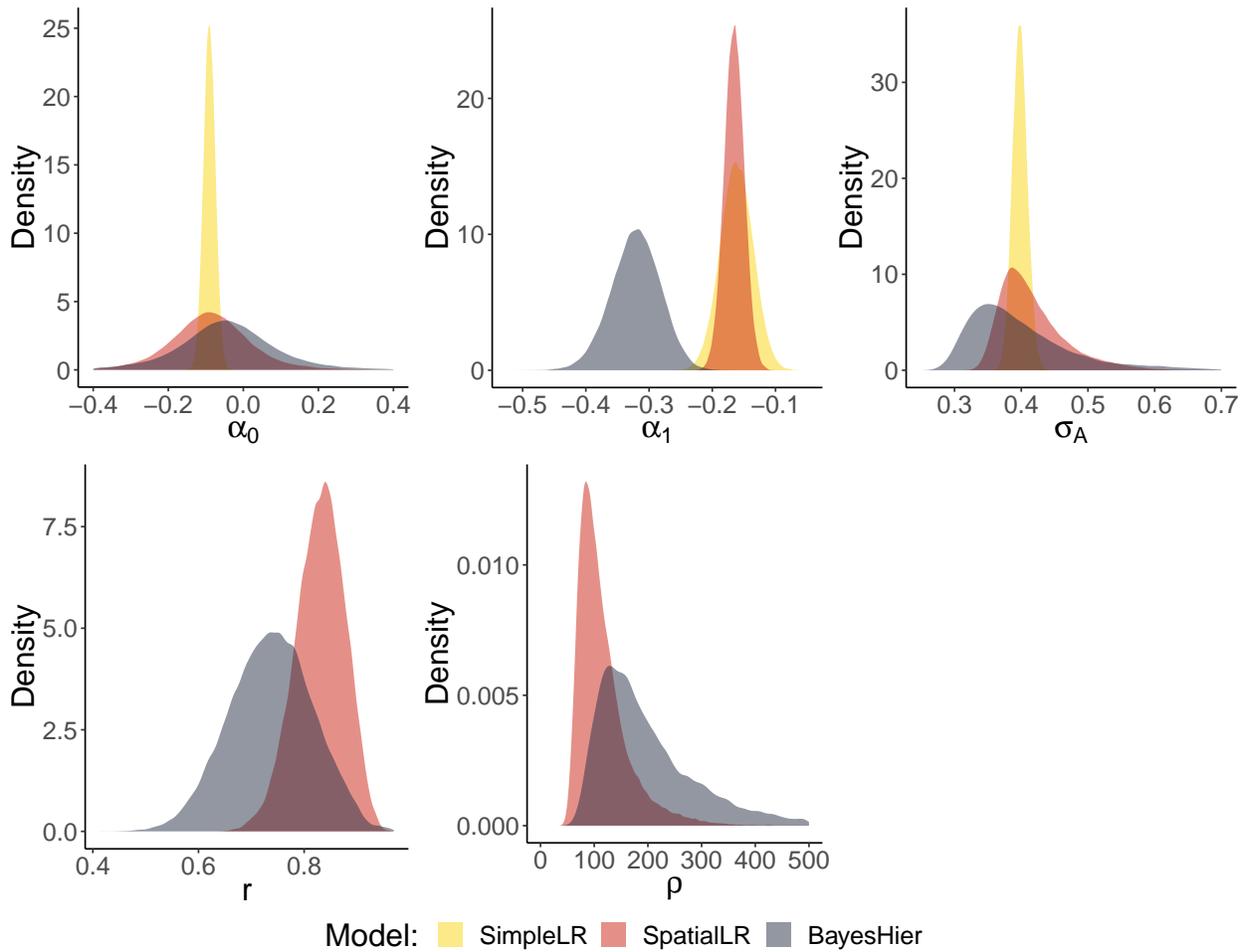}
 \label{fig:density_nothin}
\end{figure}

The results of the analysis are as expected based on our simulation study findings. Table \ref{table:realresults} shows that the posterior means for $\alpha_1$ in the simple and spatial linear regression models are much closer to zero than in the hierarchical model, and the variance is inflated, as we expect because of ME. The magnitude of the estimated effect is 53\% larger for the full model than for the standard models. We visualize these results in the density plots in Figure \ref{fig:density_nothin}. We also see the posterior intervals and means for the atom column locations in Table \ref{table:realresults}. The spatial linear regression model puts a wider interval on the intercept term $\alpha_0$ than the simple linear regression model, which allows for a narrower interval around the regression coefficient of interest $\alpha_1$. The spatial linear regression credible interval for $\alpha_1$ does not overlap with the interval from the hierarchical model, providing strong evidence of attenuation. 

All three models indicate strong evidence of a negative relationship between A-site column displacement and B-site intensity through the parameter $\alpha_1$. In other words, A-site column locations tend to be further from B-sites with higher proportions of magnesium. These findings are consistent with observations made with x-ray diffraction that propose the distribution of magnesium and niobium directly influences the bonding between lead and oxygen \citep{Chen1996,Jeong2005a}. In addition to what we present here, we have found overwhelming evidence $\alpha_1 \neq 0$ using Bayes Factors through stochastic search variable selection. However, marginal likelihoods in this setup are notoriously sensitive to untestable model assumptions \citep{gelman2013bayesian}, so we relegate these results to the Supplementary Information.

When interpreting these findings, we need to be careful that spatial confounding is not biasing our estimate of $\alpha_1$ \citep{hodges2010adding, paciorek2010importance}. Spatial confounding is most prevalent when the covariates have strong spatial dependence, but in our exploratory data analysis, we found no evidence of spatial correlation in our covariate and no evidence of correlation between the covariate and the residuals of OLS estimates. Furthermore, the posterior means of $\alpha_1$ for the simple and spatial linear regression models are equal and the posterior standard deviation for the spatial linear regression model is less than that of the simple linear regression model. Finally, our results align with our simulation study. This leads us to conclude that the difference in the posterior distributions of $\alpha_1$ for the hierarchical model compared to the simple and spatial linear regression models is due to measurement error, not confounding. 

\begin{figure}
 \centering
  \caption{95\% posterior regions (circles) and means (points) for atom column locations from the inset image in Figure \ref{fig:STEMimage}.}
 \includegraphics[page=1, width = 1\textwidth]{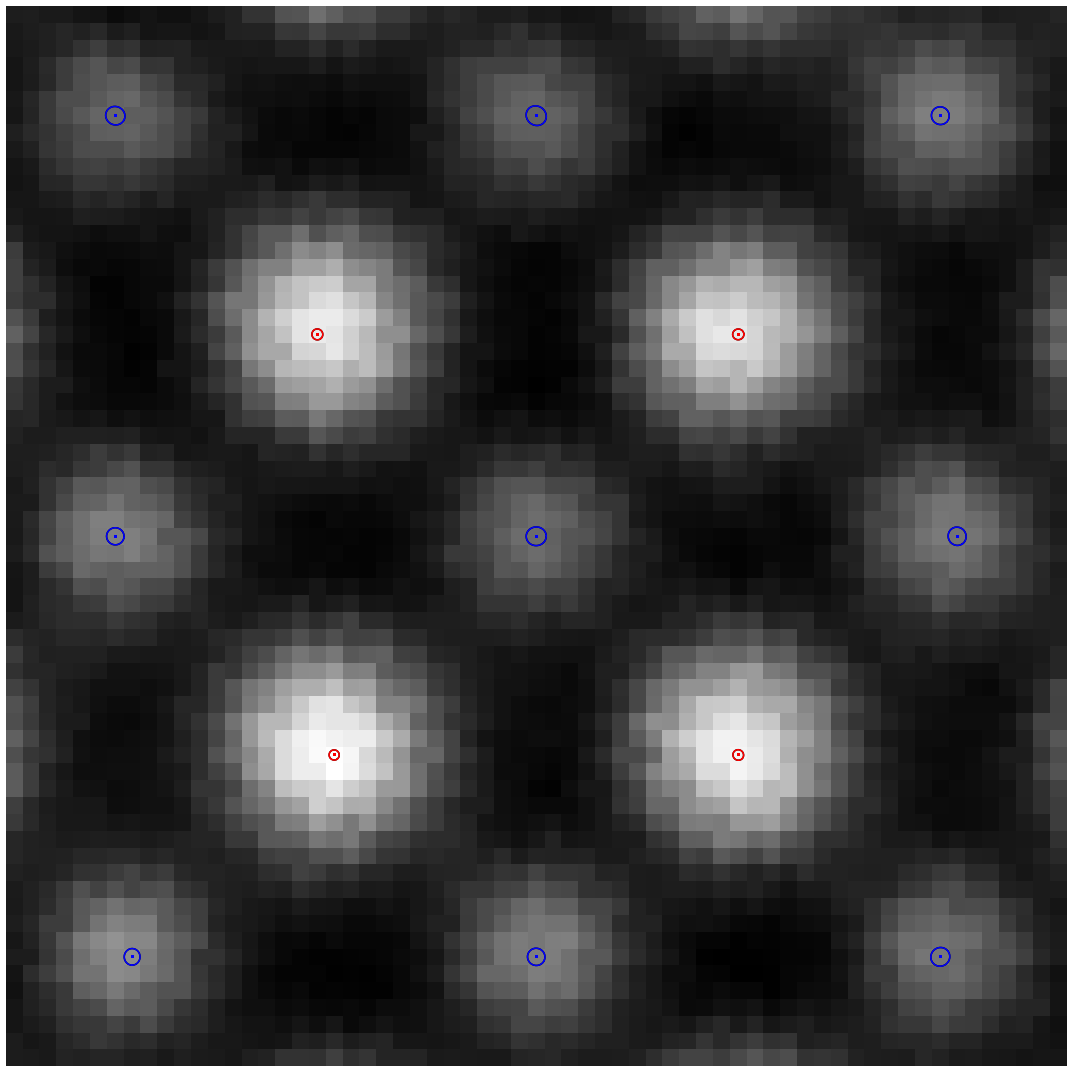}
 \label{fig:ellipses}
\end{figure}

\begin{table}[]
\begin{center}
\caption{Posterior mean and highest posterior density 95\% credible intervals for the 5 common parameters among the hierarchical, spatial linear regression, and spatial linear regression models.}
\begin{tabular}{c|rrrrrr}
\multicolumn{1}{l}{} & \multicolumn{2}{c}{Hierarchical Model}    & \multicolumn{2}{c}{Spatial LR}   & \multicolumn{2}{c}{Simple LR}   \\
\multicolumn{1}{l}{} & \multicolumn{1}{l}{Mean} & \multicolumn{1}{l}{Credible Int.} & \multicolumn{1}{l}{Mean} & \multicolumn{1}{l}{Credible Int.} & \multicolumn{1}{l}{Mean} & \multicolumn{1}{l}{Credible Int.} \\
\hline
$\alpha_0$ & -0.06 & (-0.34, 0.21) & -0.09 & (-0.32, 0.15) & -0.09 & (-0.12, -0.06)\\
$\alpha_1$ & -0.29 & (-0.36, -0.23) & -0.19 & (-0.22, -0.16) & -0.19 & (-0.24, -0.13)\\
$\sigma_A$ & 0.38 & (0.28, 0.53) & 0.42 & (0.34, 0.52) & 0.40 & (0.37, 0.42)\\
$r$ & 0.72 & (0.56, 0.87) & 0.83 & (0.73, 0.91) & $-$ & $-$\\
$\rho$ & 205 & (58, 440) & 122 & (54, 225) & $-$ & $-$   
\end{tabular}
\label{table:realresults}
\end{center}
\end{table}

\section{Discussion}
Electron microscopy imaging techniques will continue to improve and provide us with an ever clearer picture of how local physical and chemical differences contribute to global material properties. This article describes a spatial Bayesian hierarchical model that accounts for ME in locations for atomic-scale images of crystalline materials. Our new method is a dramatic improvement over the standard analysis techniques, and as such we hope it will become an impactful tool for materials scientists. We apply this model to real and simulated STEM images of PMN, and show that it outperforms spatial and simple linear regression where the estimated locations are treated as the truth. We find a negative relationship between the displacement of lead atom columns and the weighted intensity of neighboring magnesium/niobium columns, which corresponds to the proportion of niobium in those columns. The magnitude of the parameter associated with this relationship is 53\% larger in our model compared to the non-ME models, which along with our simulation study strongly suggests attenuation of the parameter in the non-ME models.

This method is computationally intensive compared to the naive models, as the number of parameters scale with the number of atom columns and the data layer uses intensities at each pixel as responses. However, using independent non-contiguous blocks around the atom columns allows the time to scale linearly with the number of columns. For the type of data explored in our application, the non-contiguous block method is limited by the maximum size of the windows around the atom columns. The blocks cannot overlap, because the information in the overlapping region would be counted twice. Rotating the image so that the angle of the rows of atom columns aligns with the blocks will help maximize the block size.  We can also modify this model to apply it to different types of crystal structures and zone axes.

\section*{Acknowledgement}
This material is based upon work supported by the National Science Foundation under Grant No. DGE-1633587. MJC, ECD, and JML gratefully acknowledge support for this work from the National Science Foundation, as part of the Center for Dielectrics and Piezoelectric under Grant Nos. IIP-1361571 and IIP-1361503.  This work was performed in part at the Analytical Instrumentation Facility (AIF) at North Carolina State University, which is supported by the State of North Carolina and the National Science Foundation (ECCS-1542015). AIF is a member of the North Carolina Research Triangle Nanotechnology Network (RTNN), a site in the National Nanotechnology Coordinated Infrastructure (NNCI). 

\section*{Appendix}
\appendix
\section{Finding Initial Atom Column Locations}
We adopt the methods described by \cite{sang2014atom} to find the initial estimates of the atom column locations, first by using the NCC to find the region for each column and then using the intensity-weighted average of the pixels as an initial location estimate. We use nonlinear least squares to fit the equation
%\begin{equation}
\begin{multline}
g(x,y) = A\exp 
 \Bigg\{
 -\Big [\frac{\big((x-x_0)\cos\theta+(y-y_0)\sin\theta\big)^2}{\sigma_1^2}\\
 -\frac{\big((x-x_0)\sin\theta-(y-y_0)\cos\theta \big)^2}{\sigma_2^2}\Big ]
 \Bigg\}
+Z,
\label{equation:gauss_fit} 
\end{multline}
%\end{equation}
where $g(x,y)$ is the atom column intensity, $Z$ is the background intensity at pixel $(x,y)$, $A$ is the peak intensity with background removed, $\theta$ is the rotation angle, and $\sigma_1$ and $\sigma_2$ are the standard deviations along the axes of the ellipse. The initial value for background intensity $Z_0$ is the difference in the median and standard deviation of the column intensity, and the initial amplitude estimate $A_0$ is the difference between $Z_0$ and the maximum intensity for the column. The coordinates $(x_0, y_0)$ are the true atom column position. We use non-linear least squares to minimize (over $\{A, \theta, \sigma_1^2, \sigma_2^2, z, x_0, y_0\}$) the average squared error between $g(x,y)$ and the fitted model and obtain estimates of $x_0$ and $y_0$. The spatial and simple linear regression models described in Section 3 used these fitted values as the known atom column locations and intensities, and the hierarchical model uses them as initial values for the MCMC algorithm used to sample from the posterior.

\begin{singlespace}
	\bibliographystyle{asa}
	\bibliography{refs}
\end{singlespace}

\newpage
%%%%%%%%%% Prefix a "S" to all equations, figures, tables and reset the counter %%%%%%%%%%
\setcounter{equation}{0}
\setcounter{figure}{0}
\setcounter{table}{0}
\setcounter{page}{1}
\setcounter{section}{0}
\makeatletter
\renewcommand{\theequation}{S\arabic{equation}}
\renewcommand{\thefigure}{S\arabic{figure}}
\renewcommand{\thesection}{S\arabic{section}}
\renewcommand{\thetable}{S\arabic{table}}
\renewcommand{\bibnumfmt}[1]{[S#1]}
\renewcommand{\citenumfont}[1]{S#1}
%%%%%%%%%% Prefix a "S" to all equations, figures, tables and reset the counter %%%%%%%%%%
\nolinenumbers
\resetlinenumber
\title{Supplement to ``Accounting for Location Measurement Error in Imaging Data with Application to Atomic Resolution Images of Crystalline Materials"}
\author{Matthew J. Miller$^a$, Matthew J. Cabral$^b$, Elizabeth C. Dickey$^b$, \\James M. LeBeau$^c$,  and Brian J. Reich$^a$}
\date{}

\maketitle
\begin{singlespace}
\vspace{-.5in}
\begin{enumerate}
\item[$^a$]  Department of Statistics, North Carolina State University, Raleigh, NC
\item[$^b$]  Department of Materials Science and Engineering, North Carolina State University Raleigh, NC
\item[$^c$]  Department of Materials Science and Engineering, Massachusetts Institute of Technology, Cambridge, MA
\end{enumerate}
\end{singlespace}
%\linenumbers
\doublespacing
\section{Justification for Model Choices}
\subsection{Gaussian Kernel for Intensities}
\begin{figure}
    \centering
    \caption{Horizontal and vertical traces of the A-site and B-site columns from Figure \ref{fig:STEMimage}. The gray curves are from the original columns, the black curves are the means and the red curves are Gaussian fits of the means.}
    \includegraphics[width=1\textwidth]{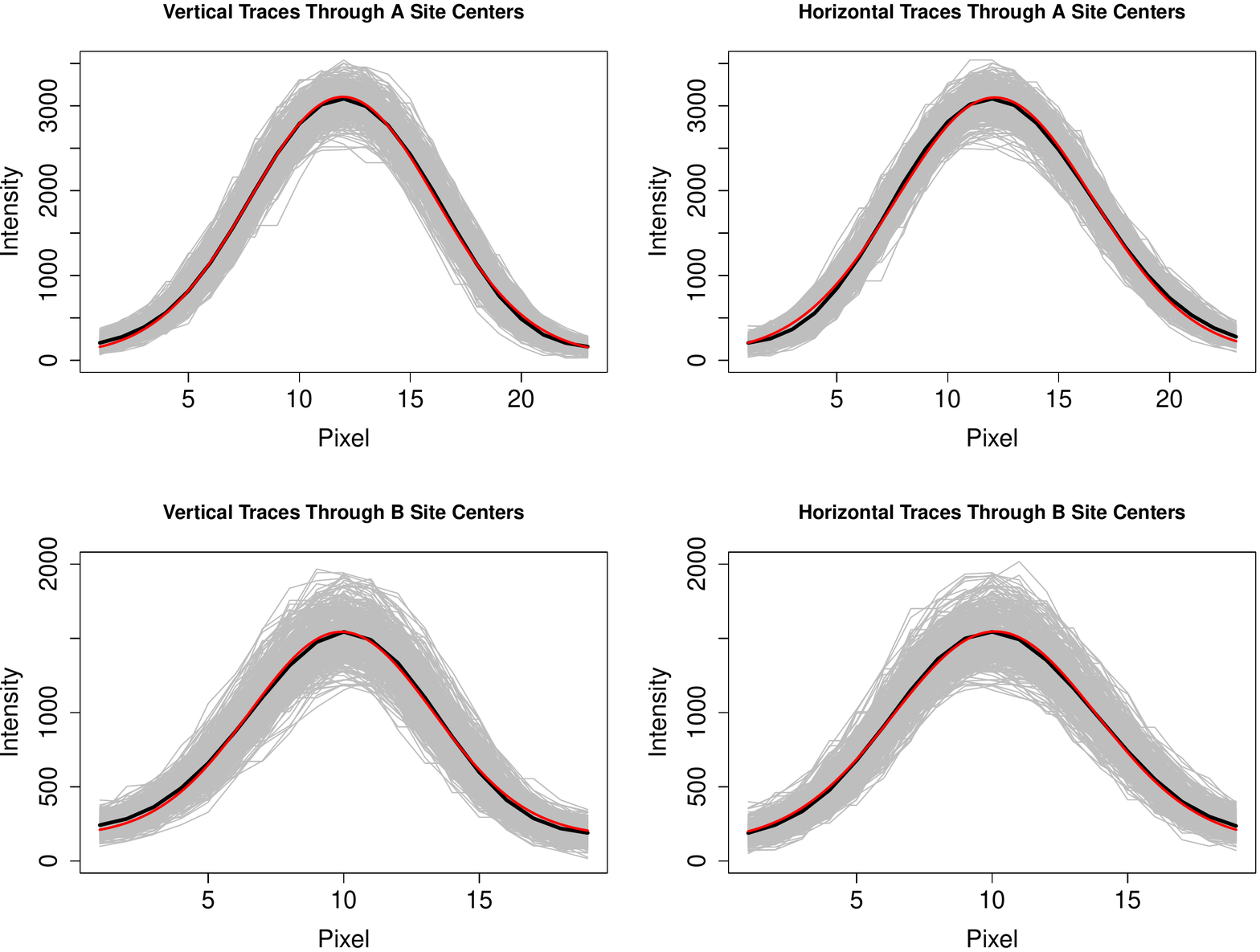}
    \label{fig:AtomTraces}
\end{figure}

In (\ref{equation:data_layer}) we assume a Gaussian kernel for the intensity decay moving away from an atom column location.  To verity this kernel is appropriate, we fit Gaussian curves to the means of the horizontal and vertical traces of the atom column pixels in Figure \ref{fig:STEMimage}. We compare these fitted curves to the mean traces in Figure \ref{fig:AtomTraces} and they align nearly perfectly. The correlation between the mean and fitted curves is at least $0.999$ for each atom column type and direction. Therefore we believe Gaussian kernels are appropriate for the model.

\subsection{Exponential Covariance Functions}
\begin{figure}
    \centering
    \caption{Fitted empirical variogram of residuals of OLS estimates on A- and B-site column intensities using the exponential covariance function.}
    \includegraphics[width = 1\textwidth]{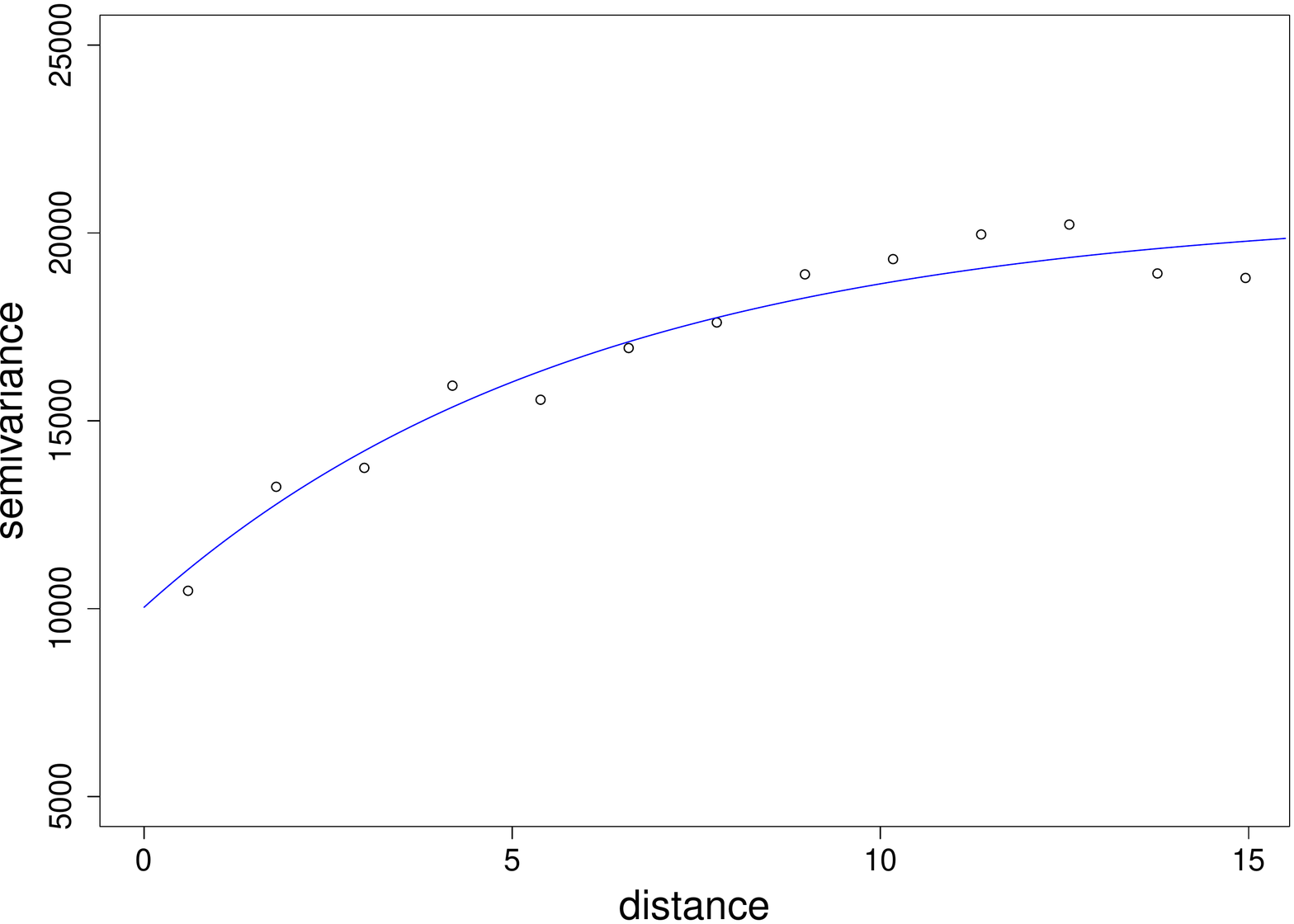}
    \label{fig:variointensities}
\end{figure}

\begin{figure}
    \centering
    \caption{Fitted empirical semivariogram of the residuals of OLS estimates on x- and y- displacements of A-site atom columns using the exponential covariance function.}
    \includegraphics[width = 1\textwidth]{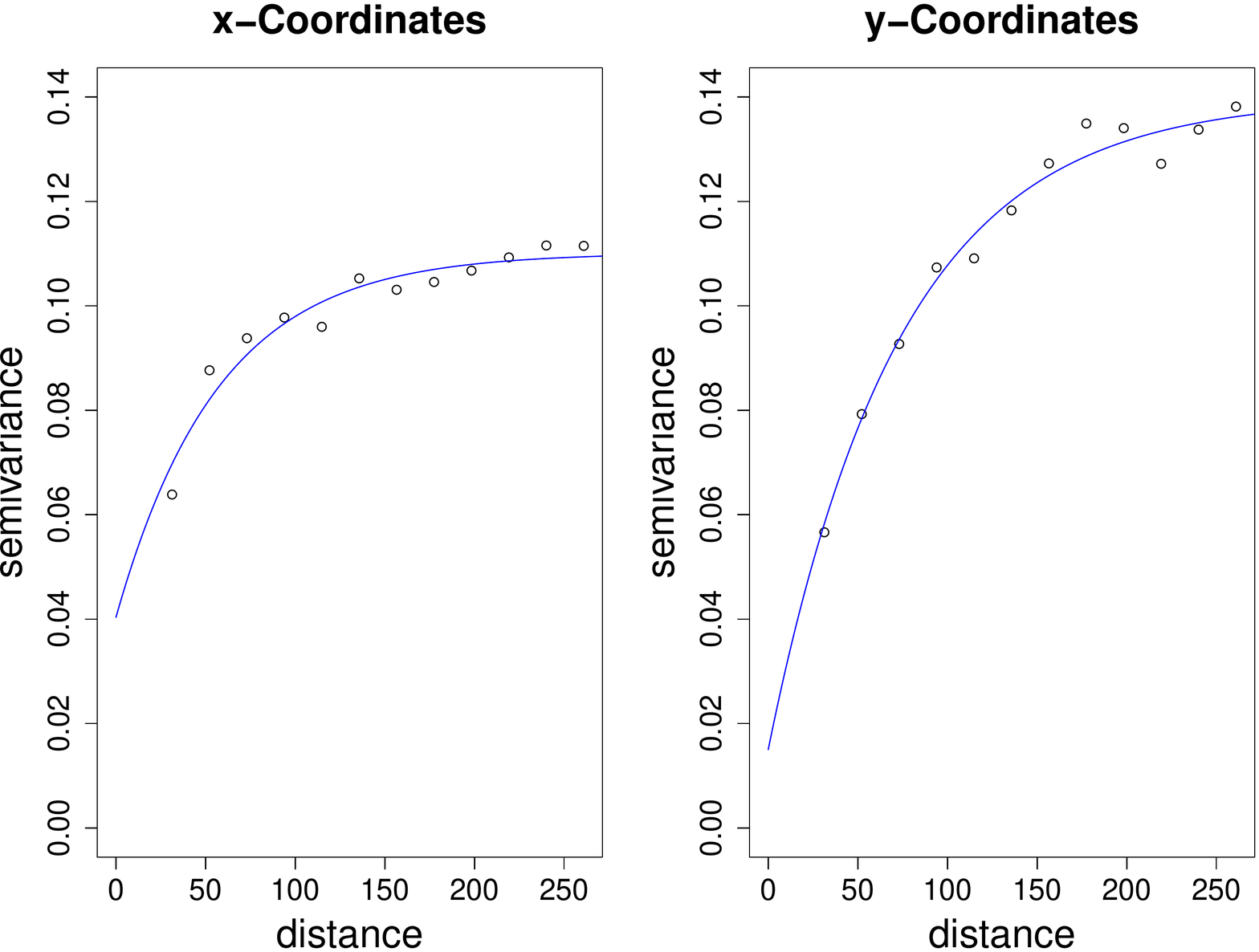}
    \label{fig:varioxy}
\end{figure}

We assume an exponential correlation function in (\ref{equation:cov}). In addition to exponential covariance functions being in the desirable Mat\`{e}rn class, they are relatively straightforward to incorporate into an MCMC sampling framework. We justify their use in our model by fitting exponential covariance functions to empirical variograms of the residuals of OLS estimates. In the data layer, we fix bandwidths $\psi_A = 5$ and $\psi_B=4$ and examine the residuals of the OLS estimates from equation (\ref{equation:appx_datalayer}) based on the data from Figure \ref{fig:STEMimage}. We then use the {\bf geoR} package \citep{ribeiro2007geor} to calculate, fit and plot the empirical variogram seen in Figure \ref{fig:variointensities}. Similarly, we fit the empirical variogram of the residuals of the OLS estimates from equation (\ref{equation:s_A}) in Figure \ref{fig:varioxy}. We see in both figures that the exponential covariance function fits the data well, and that the use of exponential covariance functions in our model is appropriate.

\section{Hypothesis Testing Using Stochastic Search Variable Selection}

\begin{figure}
    \centering
    \caption{Posterior densities in the five common parameters between hierarchical, spatial linear regression, and simple linear regression models, where stochastic search variable selection is included. The plots are almost identical to the ones seen in Figure \ref{fig:density_nothin}}
    \includegraphics[page=1,width=1\textwidth]{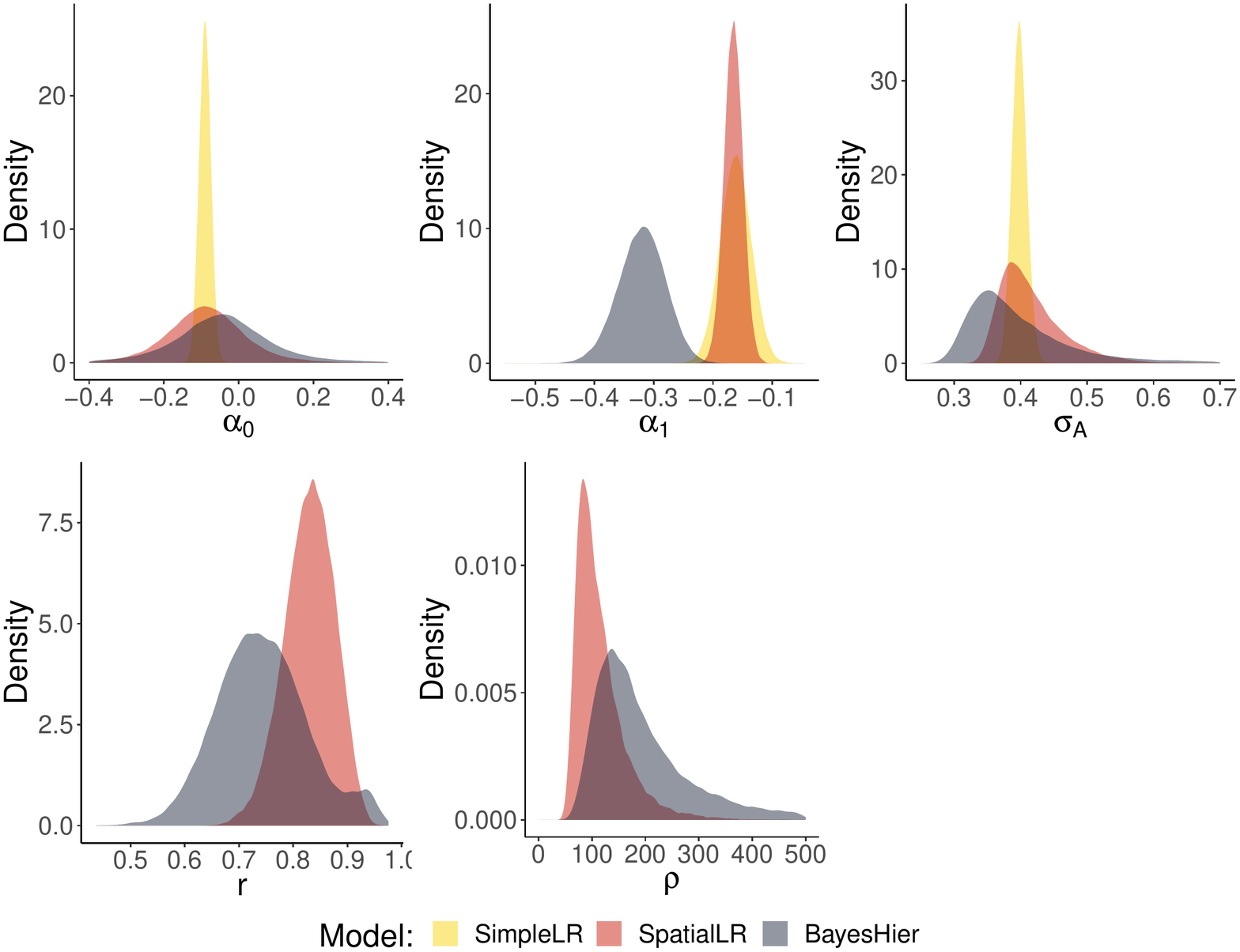}
    \label{fig:SSVS}
\end{figure}

Our analysis relied on inspecting the posterior interval of $\alpha_1$ to test whether it was zero.  Alternatively, we can introduce a Bayes factor hypothesis testing approach into our models using stochastic search variable selection. Let slope parameter $\alpha_1 = \gamma\eta$, where $\gamma \sim N(0, 10^2)$  and $\eta \sim \text{Bernoulli}(0.5)$. If $\eta = 0$, then $\alpha_1 = 0$ and if $\eta = 1$, $\alpha_1 \sim N(0,10^2)$. We use MCMC as before to collect 90,000 posterior samples after a 10,000 iteration burn-in. We find $\eta = 1$ for all posterior samples in each model, providing overwhelming evidence that $\alpha_1 \neq 0$. The 95\% highest posterior density intervals for $\alpha_1$ are $(-0.21, -0.11)$ for the simple linear regression model, $(-0.20, -0.13)$ for the spatial linear regression model and $(-0.40, -0.24)$ for the hierarchical model, which are similar to the intervals in Table \ref{table:realresults}, and again we see strong evidence of attenuation. We plot the posterior densities of the common parameters in Figure \ref{fig:SSVS}, and see that the plots are almost identical to the ones in Figure \ref{fig:density_nothin}.

\section{Sensitivity Analysis}
We examine the sensitivity of the hierarchical model to the choice of $a_i$ and $b_i$, the hyperparameters on the variance $\sigma^2_{\beta_i}$ of the intensity parameters $\beta_{ij}$, where $i\in \{A, B\}$. These parameters depend on OLS estimates of the intensity parameters as well as a chosen variance, so we perform a simulation study on increases and decreases of 50\% in this variance. We also examine performance when shrinking the widths of the A-site and B-site windows by two pixels. For this study, we use data generated with $\sigma = 140$ and $r_{pix} = 0.57$, the same parameters as Table \ref{table:simresults}, column 1. Table \ref{table:sensitivity} shows that increasing and decreasing the variance of these hyperparameters do not meaningfully change the coverage or the MSE of $\alpha_1$ from the original. We also see the coverage of $\alpha_1$ slightly lower with the smaller window, but that it is still within to Monte Carlo standard errors of $95 \%$.
\begin{table}[]
\caption{Sensitivity analysis results when decreasing and increasing variance on the prior of the atom column intensity variances by $50\%$ and when decreasing the width the A-sites and B-sites by 2.}
\begin{tabular}{lllll}
                          & Original        & Dec. Variance  & Inc. Variance   & Smaller Window \\
\hline
Bias                      & -0.002 (0.0016) & 0.002 (0.0015) & -0.003 (0.0016) & 0.009 (0.0015) \\
Mean Post. SD             & 0.017 (0.0002)  & 0.016 (0.0001) & 0.017 (0.0002)  & 0.016 (0.0002) \\
Coverage                  & 95 (2.2)        & 97 (1.7)       & 96 (2.0)        & 92 (2.7)       \\
$\widehat{MSE}\times 100$ & 0.03 (0.004)    & 0.02 (0.003)   & 0.03 (0.003)    & 0.03 (0.004)  
\end{tabular}
\label{table:sensitivity}
\end{table}

\section{MCMC Derivations}
\noindent Here we provide the derivations of the full conditional derivations used for MCMC. Let $N_i$ be the number of atoms of type $i$ and $M_{ij}$ be the the number of pixels in window $\mathbf{W}_{ij}$.
\subsection{Data Layer}
\subsubsection{Background Intensity Parameter $\beta_0$}
Let the prior distribution for $\beta_0$ be $N(0, \sigma^2_{\beta_0}$), and $\mathbf{Q}_i$ be the pixel-pixel precision matrix for a box around atom columns of type $i$. Let $\mathbf{X}_{ij} = [\exp(-\frac{\lVert \mathbf{p}_{ij1} - \mathbf{s}_ij \rVert^2}{2\psi_i^2}), \dots, \exp(-\frac{\lVert \mathbf{p}_{ijN_i} - \mathbf{s}_ij \rVert^2}{2\psi_i^2})]^T$ and $\mathbf{u}_{ij} = \beta_0\mathbf{1}_{M_i} + \beta_{ij}\mathbf{X}_{ij}$. Then, using the approximate likelihood from equation \ref{equation:approx_like},
\begin{align*}
p(\beta_0&|\mathbf{Y},\mathbf{\Theta}_{ [-\beta_0]}) \propto p(\mathbf{Y}|\mathbf{\Theta})p(\beta_0) \\
&\approx \prod_{i\in \{A,B\}}\prod_{j=1}^{N_i}p(\mathbf{Y}_{ij}|\mathbf{s}_{ij},\mathbf{\Theta})\exp{(-\frac{\beta_0^2}{2\sigma_{\beta_0}^2})} \\
&\propto \exp{(-\frac{\beta_0^2}{2\sigma_{\beta_0}^2})} \prod_{i\in \{A,B\}}\prod_{j=1}^{N_i}\exp{\big [-\frac{1}{2}(\mathbf{Y}_{ij} - \mathbf{\mu}_{ij})^T\mathbf{Q}_i(\mathbf{Y}_{ij} - \mathbf{\mu}_{ij})\big ]} \\
&\propto \exp{(-\frac{\beta_0^2}{2\sigma_{\beta_0}^2})}  \\
&\quad \times \prod_{i\in \{A,B\}}\prod_{j=1}^{N_i}\exp{\Big\{-\frac{1}{2}\big [\beta_0^2\mathbf{1}_{M_i}^T\mathbf{Q}_i\mathbf{1}_{M_i}} - 2\beta_0(\mathbf{Y}_{ij}-\beta_{ij}\mathbf{X}_{ij})^T\mathbf{Q}_i\mathbf{1}_{M_i}\big ]\Big\} \\
&\propto \exp{\Big\{-\frac{1}{2}\big [\beta_0^2(\frac{1}{\sigma_{\beta_0}^2} + \sum_{i \in \{A,B\}}N_i(\mathbf{1}_{M_i}^T\mathbf{Q}_i\mathbf{1}_{M_i}))} \\
&\quad - 2\beta_0\sum_{i \in \{A,B\}}\sum_{j=1}^{N_i}(\mathbf{Y}_{ij}-\beta_{ij}\mathbf{X}_{ij})^T\mathbf{Q}_i\mathbf{1}_{M_i}\big ] \Big\} \\
&\propto \exp{\Big\{-\frac{1}{2}\big [\frac{1}{\sigma_{\beta_0}^2} + \sum_{i \in \{A,B\}}N_i(\mathbf{1}_{M_i}^T\mathbf{Q}_i\mathbf{1}_{M_i})\big ]} \\
&\quad \times \big [\beta_0^2 - 2\beta_0(\frac{1}{\sigma_{\beta_0}^2} + \sum_{i \in \{A,B\}}N_i(\mathbf{1}_{M_i}^T\mathbf{Q}_i\mathbf{1}_{M_i}))^{-1} \\
&\quad \times \sum_{j=1}^{N_i}(\mathbf{Y}_{ij}-\beta_{ij}\mathbf{X}_{ij})^T\mathbf{Q}_i\mathbf{1}_{M_i}\big ]\Big\}. \\
\end{align*}
If we let $V_{\beta_0} = \frac{1}{\sigma_{\beta_0}^2} + \sum_{i \in \{A,B\}}N_i(\mathbf{1}_{M_i}^T\mathbf{Q}_i\mathbf{1}_{M_i})$ and $M_{\beta_0} = \sum_{j=1}^{N_i}(\mathbf{Y}_{ij}-\beta_{ij}\mathbf{X}_{ij})^T\mathbf{Q}_i\mathbf{1}_{M_i}$, then 
\begin{equation}
 \beta_0|\mathbf{Y},\mathbf{\Theta}_{ [-\beta_0]} \sim N(\frac{M_{\beta_0}}{V_{\beta_0}}, \frac{1}{V_{\beta_0}}).
\label{equation:beta0_FC}
\end{equation}

\subsubsection{A-site Intensity Parameters $\beta_{Aj}$}
Let the prior distribution for $\beta_{Aj}$ be $N(\mu_{\beta_A}, \sigma^2_\beta)$, and let $\mu_{Aj}$, $\mathbf{X}_{Aj}$ and $\mathbf{Q}_A$ be defined as in the previous subsection. Then,

\begin{align*}
p(\beta_{Aj}&|\mathbf{Y},\mathbf{\Theta}_{ [-\beta_{Aj}]}) \propto p(\mathbf{Y}|\mathbf{\Theta})p(\beta_{Aj}) &\\
&\approx \prod_{i\in \{A,B\}}\prod_{j=1}^{N_i}p(\mathbf{Y}_{ij}|\mathbf{s}_{ij},\mathbf{\Theta})\exp(-\frac{(\beta_{Aj}-\mu_{\beta_A})^2}{2\sigma_{\beta_A}^2}) &\\
&\propto p(\mathbf{Y}_{Aj}|s_{Aj},\mathbf{\Theta})\exp(-\frac{(\beta_{Aj}-\mu_{\beta_A})^2}{2\sigma_{\beta_A}^2}) &\\
&\propto \exp\Big\{-\frac{1}{2}\big [(\mathbf{Y}_{Aj}-\mu_{Aj})^T\mathbf{Q}_A(\mathbf{Y}_{Aj}-\mu_{Aj}) + (\beta_{Aj}-\mu_{\beta_A})^2\big ]\Big\} &\\
&\propto\exp\Big\{-\frac{1}{2}\big [\beta_{Aj}^2(\mathbf{X}_{Aj}^T\mathbf{Q}_A\mathbf{X}_{Aj} + \frac{1}{\sigma^2_{\beta_A}}) &\\
& \quad -2\beta_{Aj}((\mathbf{Y}_{Aj}-\beta_0\mathbf{1}_{M_A})^T\mathbf{Q}_A(\mathbf{Y}_{Aj}-\beta_0\mathbf{1}_{M_A}) + \frac{\mu_{\beta_A}}{\sigma_\beta^2})\big ]\Big\} &
\end{align*}
Let $M_{\beta_{Aj}} = (\mathbf{Y}_{Aj}-\beta_0\mathbf{1}_{M_A})^T\mathbf{Q}_A(\mathbf{Y}_{Aj}-\beta_0\mathbf{1}_{M_A}) + \frac{\mu_{\beta_A}}{\sigma_\beta^2}$ and $V_{\beta_{Aj}} = \mathbf{X}_{Aj}^T\mathbf{Q}_A\mathbf{X}_{Aj} + \frac{1}{\sigma^2_{\beta_A}}$. After factoring out $V_{\beta_{Aj}}$ and completing the square, we see 
\begin{equation}
 \beta_{Aj}|\mathbf{Y},\mathbf{\Theta}_{ [-\beta_{Aj}]} \sim N(\frac{M_{\beta_{Aj}}}{V_{\beta_{Aj}}}, \frac{1}{V_{\beta_{Aj}}})
\label{equation:betaAj_FC} 
\end{equation}
\subsection{B-site Intensity Parameters $\beta_{Bj}$}
Because the B-sites are used in determining the A-site locations, we cannot derive a full conditional, and must use Metropolis sampling instead. However, we draw from a $N(\frac{M_{\beta_{Bj}}}{V_{\beta_{Bj}}}, \frac{1}{V_{\beta_{Bj}}})$ to get our candidate instead of using the standard method. $M_{\beta_{Bj}}$ and $M_{\beta_{Bj}}$ are calculated the same way as $M_{\beta_{Aj}}$ and $M_{\beta_{Aj}}$, but with replacing the $A$'s for $B$'s.

\subsubsection{Variance Parameter $\sigma^2$}
Let the prior distribution for $\sigma^2$ be $\text{InvGamma}(c,d)$. Let $\mu_{ij}$ be defined as before and $\mathbf{Q}_i^* = \sigma^2\mathbf{Q}_i$. Then,
\begin{align*}
 p(\sigma^2&|\mathbf{Y},\mathbf{\Theta}_{ [-\sigma^2]}) \propto p(\mathbf{Y}|\mathbf{\Theta})p(\sigma^2) &\\
&\approx \prod_{i\in \{A,B\}}\prod_{j=1}^{N_i}p(\mathbf{Y}_{ij}|\mathbf{s}_{ij},\mathbf{\Theta})(\sigma^2)^{-c-1}\exp(-\frac{d}{\sigma^2}) &\\
&\propto (\sigma^2)^{-(\frac{N_a+N_b}{2}+c)-1}\exp\Big\{-\frac{1}{2\sigma^2}\big [2d &\\
&\quad + \sum_{i \in \{A,B\}}\sum_{j = 1}^{N_i}(\mathbf{Y}_{ij}-\mu_{ij})^T\mathbf{Q}_i^*(\mathbf{Y}_{ij}-\mu_{ij})\big ]\Big\},
\end{align*}
which is the kernel of an inverse gamma distribution. So,
\begin{flalign}
 \sigma^2|\mathbf{Y},\mathbf{\Theta}_{ [-\sigma^2]} \sim \text{IG}(\frac{N_a+N_b}{2}+c, d + \frac{1}{2}\sum_{i \in \{A,B\}}\sum_{j = 1}^{N_i}(\mathbf{Y}_{ij}-\mu_{ij})^T\mathbf{Q}_i^*(\mathbf{Y}_{ij}-\mu_{ij}) )
\end{flalign}
\subsection{Process Layer}
Let $\mathbf{\delta}_x = (\delta_{x1}, \delta_{x2}, \cdots, \delta_{xa})^T$, where $\delta_{xj} = s_{A_jx}-U_{A_jx}$, the difference of the x-coordinates of the $\text{j}^{th}$ atom column location and the $2\times 1$ vector defined in equation \ref{equation:unweighted}. Let $\mathbf{\Psi}_x = (\Psi_{x1}, \Psi_{x2}, \cdots, \Psi_{xa})^T$, where $\Psi_{xj} = W_{A_jx} - U_{A_jx}$, the difference of the x-coordinates defined in equations $\ref{equation:weighted}$ and $\ref{equation:unweighted}$, respectively. Define $\mathbf{\delta}_y$ and $\mathbf{\Psi}_y$ similarly. Then, the distributions of their likelihoods are:
\begin{equation}
 \mathbf{\delta}_i|\mathbf{\Theta} \overset{\text{ind}}{\sim} N(\alpha_0\mathbf{1}_a + \alpha_1\mathbf{\Psi}_i, \mathbf{V}),
\label{equation: process_like}
\end{equation}
where
\begin{equation}
 \mathbf{V} = \sigma^2_A\big [(1-r)\mathbf{I}_a + r\exp{(-\frac{1}{\rho}\mathbf{D}_A})\big ].
\end{equation} 
$\mathbf{\Theta} = (\alpha_0, \alpha_1, \sigma_a, r, \rho, \mathbf{S}_B, \beta_{B1}, \dots, \beta_{Bb})^T$, where $\mathbf{S}_B$ is the $b\times 2$ matrix of B-site locations and $b$ is the number of B-sites. $\mathbf{1}_a$ is an $a\times 1$ vector of 1's, $\mathbf{D}_A$ is the Euclidean distance matrix between the A-sites and $\mathbf{I}_a$ is the $a\times a$ identity matrix, where $a$ is the number of A-sites. Finally, let $\mathbf{\Theta}_{ [-p]}$ be the vector $\Theta$ with parameter $p$ removed.

\subsubsection{Intercept Parameter $\alpha_0$}
Recall from \ref{table:parameters} the prior distribution for $\alpha_0$ is $N(0,1000^2)$. We will generalize here and let $\alpha_0 \sim N(0, \sigma_{\alpha_0}^2)$. Let $\mathbf{\mu}_i = \mathbf{\delta}_i -\alpha_0\mathbf{1}_a -\alpha_1\mathbf{\Psi}_i$ Then, we derive the full conditional distribution: 
\begin{align*}
p(\alpha_0|\mathbf{\delta}_x, \mathbf{\delta}_y, \mathbf{\Theta}_{ [-\alpha_0]}) &\propto p(\mathbf{\delta}_x|\mathbf{\Theta})p(\mathbf{\delta}_y|\mathbf{\Theta})p(\alpha_0)\\ 
&\propto \exp\Big\{-\frac{1}{2}\big [\sum_{i\in \{x,y\}}(\mathbf{\mu}_i^T\mathbf{V}^{-1}\mathbf{\mu}_i)+ \frac{\alpha_0^2}{\sigma_{\alpha_0}^2}\big ]\Big\}\\
&\propto \exp\Big\{-\frac{1}{2}\big [\sum_{i\in\{x,y\}}(\alpha_0^2\mathbf{1}_a^T\mathbf{V}^{-1}\mathbf{1}_a\\ 
&\quad - 2\alpha_0(\mathbf{\delta}_i-\alpha_1\mathbf{\Psi}_i)^T\mathbf{V}^{-1}\mathbf{1}_a) + \frac{\alpha_0^2}{\sigma_{\alpha_0}^2}\big ]\Big\}\\
&\propto \exp\Big\{-\frac{1}{2}\big [\alpha_0^2(2(\mathbf{1}_a^T\mathbf{V}^{-1}\mathbf{1}_a) + \frac{1}{\sigma^2_{\alpha_0}})\\
&\quad - 2\alpha_0\sum_{i \in\{x,y\}}(\mathbf{\delta}_i-\alpha_1\mathbf{\Psi}_i)\mathbf{V}^{-1}\mathbf{1}_a\big ]\Big\}\\
&\propto \exp\Big\{-\frac{2(\mathbf{1}_a^T\mathbf{V}^{-1}\mathbf{1}_a) + \frac{1}{\sigma^2_{\alpha_0}}}{2}\big [\alpha_0^2 \\
&\quad - 2\alpha_0\frac{\sum_{i \in\{x,y\}}(\mathbf{\delta}_i-\alpha_1\mathbf{\Psi}_i)\mathbf{V}^{-1}\mathbf{1}_a}{2(\mathbf{1}_a^T\mathbf{V}^{-1}\mathbf{1}_a) + \frac{1}{\sigma^2_{\alpha_0}}}\big ]\Big\}.
\end{align*}
If we let $V_{\alpha_0} = 2(\mathbf{1}_a^T\mathbf{V}^{-1}\mathbf{1}_a) + \frac{1}{\sigma^2_{\alpha_0}}$ and $M_{\alpha_0} = \sum_{i \in\{x,y\}}(\mathbf{\delta}_i-\alpha_1\mathbf{\Psi}_i)\mathbf{V}^{-1}\mathbf{1}_a$, we see after completing the square that 
\begin{equation}
 \alpha_0|\mathbf{\delta}_x, \mathbf{\delta}_y, \mathbf{\Theta}_{ [-\alpha_0]} \sim N(\frac{M_{\alpha_0}}{V_{\alpha_0}}, \frac{1}{V_{\alpha_0}}).
\label{equation: alpha0_FC}
\end{equation}

\subsubsection{Slope Parameter $\alpha_1$}
As we did with $\alpha_0$, we will generalize the prior for $\alpha_1$ and let $\alpha_1 \sim N(0, \sigma_{\alpha_1}^2)$. Again let $\mathbf{\mu}_i$ be defined as above. Then,
\begin{align*}
 p(\alpha_1|\mathbf{\delta}_x, \mathbf{\delta}_y, \mathbf{\Theta}_{ [-\alpha_1]}) &\propto p(\mathbf{\delta}_x|\mathbf{\Theta})p(\mathbf{\delta}_y|\mathbf{\Theta})p(\alpha_1)\\ 
&\propto \exp\Big\{-\frac{1}{2}\big [\sum_{i\in \{x,y\}}(\mathbf{\mu}_i^T\mathbf{V}^{-1}\mathbf{\mu}_i)+ \frac{\alpha_1^2}{\sigma_{\alpha_1}^2}\big ]\Big\}\\
&\propto \exp\Big\{-\frac{1}{2}\big [\sum_{i\in\{x,y\}}(\alpha_1^2\mathbf{\Psi}_i \mathbf{V}^{-1} \mathbf{\Psi}_i - 2\alpha_1(\mathbf{\delta}_i-\alpha_0\mathbf{1}_a)) + \frac{\alpha_1^2}{\sigma_{\alpha_1}}\big ]\Big\}\\
&\propto \exp\Big\{-\frac{\sum_{i\in\{x,y\}}\mathbf{\Psi}_i \mathbf{V}^{-1} \mathbf{\Psi}_i+\frac{1}{\sigma^2_{\alpha_1}}}{2}\big [\alpha_1^2\\
&\quad - 2\alpha_1\frac{\sum_{i\in\{x,y\}}(\mathbf{\delta}_i-\alpha_0\mathbf{1}_a)}{\sum_{i\in\{x,y\}}\mathbf{\Psi}_i \mathbf{V}^{-1} \mathbf{\Psi}_i + \frac{1}{\sigma_{\alpha_1}^2}}\big ]\Big\}
\end{align*}
Letting $V_{\alpha_1} = \sum_{i\in\{x,y\}}\mathbf{\Psi}_i \mathbf{V}^{-1} \mathbf{\Psi}_i+\frac{1}{\sigma^2_{\alpha_1}}$ and $M_{\alpha_1} = \sum_{i\in\{x,y\}}(\mathbf{\delta}_i-\alpha_0\mathbf{1}_a)$, after completing the square we have 
\begin{equation}
 \alpha_1|\mathbf{\delta}_x, \mathbf{\delta}_y, \mathbf{\Theta}_{ [-\alpha_1]} \sim N(\frac{M_{\alpha_1}}{V_{\alpha_1}}, \frac{1}{M_{\alpha_1}}).
\label{equation: alpha1_FC}
\end{equation}

\subsubsection{Variance Parameter $\sigma^2_A$}
Let the prior distribution for $\sigma^2_A$ be $\text{InvGamma}(f,g)$. Let $\mathbf{V}^* = \frac{1}{\sigma^2_A}\mathbf{V}$, and $\mathbf{\mu}_i$ be defined the same as in the previous subsections. Then,
\begin{align*}
 p(\sigma^2_A|\mathbf{\delta}_x, \mathbf{\delta}_y, \mathbf{\Theta}_{ [-\sigma^2_A]}) &\propto p(\mathbf{\delta}_x|\mathbf{\Theta})p(\mathbf{\delta}_y|\mathbf{\Theta})p(\sigma^2_A) \\ 
 &\propto |\sigma^2_A\mathbf{V}^*|^{-1}\exp\big [-\frac{1}{2\sigma^2_A}\sum_{i \in \{x,y\}}\mathbf{\mu}_i^T(\mathbf{V}^*)^{-1}\mathbf{\mu}_i\big ]\\
 &\quad \times (\sigma^2_A)^{-f-1}\exp(-\frac{g}{\sigma^2_A}) \\
 &\propto (\sigma^2_A)^{-(f+N_A)-1}\exp\Big\{-\frac{1}{{\sigma^2_A}}(g + \frac{\sum_{i \in \{x,y\}}\mathbf{\mu}_i^T(\mathbf{V}^*)^{-1}\mathbf{\mu}_i}{2})\Big\},
\end{align*}

which is the kernel of an inverse gamma distribution, so
\begin{equation}
 \sigma^2_A|\mathbf{\delta}_x, \mathbf{\delta}_y, \mathbf{\Theta}_{[-\sigma^2_A]} \sim \text{InvGamma}(f+N_A, g + \frac{\sum_{i \in \{x,y\}}\mathbf{\mu}_i^T(\mathbf{V}^*)^{-1}\mathbf{\mu}_i}{2}).
\label{equation:sigma2A_FC}
\end{equation}

\subsection{Prior Layer}
\subsubsection{Mean and Variance for $\beta_A$ and $\beta_B$}
The mean and variance parameters for $\beta_A$ and $\beta_B$ have priors determined by the OLS estimates of the $\beta_{ij}'s$, as described in Table \ref{table:parameters}. Let $\hat{\mu_{\beta_i}}$ be the mean of the OLS estimates for the $\beta_{ij}$'s, and $a_i$ and $b_i$ be defined as in the Table \ref{table:parameters} caption. Then we have conjugate posteriors, with 
\begin{equation}
\mu_{\beta_i}|\mathbf{\beta}_A,\sigma_{\beta_i}^2,\hat{\mu_{\beta_i}} \sim N(\frac{\frac{\hat{\mu_{\beta_i}}}{1000^2} + \sum_{j=1}^{N_i}\beta_{ij}}{\frac{1}{1000^2}+\frac{N_i}{\sigma^2_{\beta_i}}}, \frac{1}{\frac{1}{1000^2}+\frac{N_i}{\sigma^2_{\beta_i}}}) 
\label{equation:mubeta_FC} 
\end{equation}
and
\begin{equation}
\sigma^2_{\beta_i}|\mu_{\beta_i},\mathbf{\beta}_A,a_i,b_i \sim \text{InvGamma}(a_i + N_i/2, b_i + \frac{1}{2}\sum_{j=1}^{N_i}(\beta_{ij}-\mu_{\beta_i})^2)
\label{equation:sigmabeta_FC} 
\end{equation}
\subsection{B-site variance $\sigma^2_B$}
Because the B-site locations are modeled as independent, the conjugate full conditional for $\sigma^2_B$ is standard. Let $\sigma^2_B \sim \text{InvGamma}(l,m)$. Let $\mathbf{\mu}_B$ be the matrix of grid locations described in Section 3.1.3 and $\mathbf{s}_B$ the matrix of B-site locations. Then,
\begin{equation}
 \sigma^2_B|\overset{\sim}{\mathbf{\mu}}_B, \mathbf{s}_B \sim \text{InvGamma}(N_B + l, m + \frac{1}{2}\sum_{j=1}^{N_B}\lVert \mathbf{s}_{Bj} - \mathbf{\mu}_{Bj}\rVert^2).
\end{equation}

%\begin{singlespace}
%	\bibliographystyle{asa}
%	\bibliography{refs}
%\end{singlespace}

\end{document}